\documentclass{ntnuthesis}

\usepackage{parskip}

\pdfoutput=1

\usepackage[utf8x]{inputenc}
\usepackage{pdfpages}
\usepackage{amsmath,amssymb,amstext,color,bm}
\usepackage{txfonts}
\usepackage{lmodern}
\usepackage{graphicx}
\usepackage[hang, nooneline]{subfigure}
\usepackage[activate=normal]{pdfcprot}
\usepackage{makeidx}
\makeindex

\usepackage{emptypage}
  
\usepackage{titlesec}
  
\usepackage{tocvsec2}
  
\usepackage[hang,small]{caption}
  
\newcommand{\pa}{\partial}
\newcommand{\mbf}{\mathbf}

\newcommand{\mcal}{\mathcal}
\newcommand{\tbf}{\textbf}
\newcommand{\tit}{\textit}
\newcommand{\trm}{\textrm}
\newcommand{\ep}{\,}
\def\bra#1{\mathinner{\langle{#1}|}}
\def\ket#1{\mathinner{|{#1}\rangle}}

\def\sumint{\hbox{$\sum$}\!\!\!\!\!\!\int}

\newcommand{\im}{\textit i}

\definecolor{structurecolor}{gray}{0.65}
\definecolor{chaptercolor}{gray}{0.65}
\definecolor{othercolor}{gray}{0.0}

\newcommand\contentlayout{
    \titleformat{\chapter}[display]  
    {\normalfont\sffamily\bfseries\LARGE}
    {}
    {-3.5ex}
    {\filright\Huge\textcolor{chaptercolor}{\thechapter}\LARGE\quad}
    [\vspace{1ex}%
    \titlerule]
    \titleformat{\section}[display]  
    {\normalfont\sffamily\bfseries\Large}
    {}
    {-3.5ex}
    {\filright\textcolor{othercolor}{\thesection}\Large\quad}
    []
    \titleformat{\subsection}[display]  
    {\normalfont\sffamily\bfseries\large}
    {}
    {-3.5ex}
    {\filright\textcolor{othercolor}{\thesubsection}\large\quad}
    []
    \titleformat{\subsubsection}[display]  
    {\normalfont\sffamily\bfseries\normalsize}
    {}
    {-3.5ex}
    {}
    []
    \titleformat{\paragraph}[runin]  
    {\normalfont\sffamily\bfseries\normalsize}
    {}
    {-3.5ex}
    {}
    []
}

\newcommand\introlayout{  
    \titleformat{\chapter}[display]  
    {\normalfont\sffamily\bfseries\LARGE}
    {}
    {-3.5ex}
    {}
    [\vspace{1ex}%
    \titlerule]
}    
  
\makeatletter
\renewcommand\part{%
  \if@openright
    \cleardoublepage
  \else
    \clearpage
  \fi
  \thispagestyle{empty}
  \if@twocolumn
    \onecolumn
    \@tempswatrue
  \else
    \@tempswafalse
  \fi
  \null\vfil
  \secdef\@part\@spart}
\makeatother

\let\oldtableofcontents\tableofcontents
\renewcommand{\tableofcontents}{%
	\begingroup
	\pagestyle{fancy}
	\fancyhf{}
	\fancyhead[ER]{\normalfont\sffamily\textcolor{chaptercolor}{\textbf{\nouppercase\leftmark}}}
	\fancyhead[OL]{\normalfont\sffamily\textcolor{chaptercolor}{\textbf{\nouppercase\rightmark}}}
	\fancyhead[EL,OR]{\thepage}
	\renewcommand{\headrulewidth}{0.4pt}
	\oldtableofcontents%
	\clearpage%
	\endgroup%
}

\begin{document}

  \maxtocdepth{section}

%
   
  \date{\today{}}

  \title{Model studies of\\\\the chiral and deconfinement\\\\transitions in QCD}
  \author{William Robb Naylor}

  \degreetype{Philosophiae Doctor}
  \faculty{Faculty of Natural Science and Technology}
  \department{Department of Physics}
  
  \serialnumber{2015:238}
  \isbnprinted{978-82-326-1132-4}
  \isbnelectronic{978-82-326-1133-1}

  \thispagestyle{empty}
\maketitle


  \pagenumbering{roman}

%
   
\introlayout

\cleardoublepage\addcontentsline{toc}{chapter}{\bf Preface}
\chapter*{Preface}

This thesis is submitted to the Norwegian University of Science and Technology (NTNU) as partial fulfilment of the requirements for the degree of Philosophiae Doctor. It is the result of four years research at the Department of Physics at NTNU under the supervision of Professor Jens O. Andersen.
  
This thesis consists of two parts: In the second part we present the three papers that form the backbone of the thesis. In the first part, we provide a short introduction to the research field and motivate and elucidate the papers presented in the second part. It is hoped that this will make the papers presented accessible to someone new to the field or from a related field.
\vspace{2cm}
  
  \begin{flushright}
  William Naylor
  
  Trondheim, July 2015
  \end{flushright}

\cleardoublepage\addcontentsline{toc}{chapter}{\bf To the non-physicist}
\chapter*{To the non-physicist}

The force that is responsible for holding the nucleus together and for creating protons and neutrons (via the binding of even smaller particles, called quarks) is the strong nuclear force. Of the four fundamental forces of nature, the strong force is (unsurprisingly) the strongest but also has extremely short range. Its short range arises from the fact that the force attracts to itself,\footnote{What does it mean for a force to interact with itself? Well, imagining a long ranged force like electromagnetism we are prompted to think of field lines flowing from positive charges to negative charges. If this force was attracted to its self then instead of flowing evenly to the negative charge, the lines instead would all flow toward one another, and quickly become entangled.} and thus produces very tangled configurations.

The theory of the strong nuclear force is Quantum ChromoDynamics (QCD). This theory accounts for the tangled nature of the force it describes. A result is that making calculations directly from the theory is almost impossible. Hence we follow the work of many others by instead using simplified models of QCD that approximate the theory for a certain range of temperatures and pressures.
QCD is also studied experimentally via high energy collisions, such as the experiments ongoing at the Large Hadron Collider at CERN. A byproduct of these collisions is extremely intense magnetic fields. Currently there is disagreement between model calculations (which we do) and brute force calculations of QCD using supercomputers (lattice-QCD) as to what the effect of these large magnetic fields is on QCD.

In this thesis we give one of the most complete studies of a model of QCD including a magnetic field and explore some of the mathematical details that have arisen as people search for consensus between models and lattice-QCD. In addition, we investigate some of the subtleties that arise when comparisons are made between model studies and lattice-QCD and suggest ways in which these can be accounted for.


\cleardoublepage\addcontentsline{toc}{chapter}{\bf Acknowledgements}
\chapter*{Acknowledgements}  

This thesis simply would not be possible without the help of my supervisor Jens O.\ Andersen. I thank you for all the help with physics, getting me through the academic grind and for taking on someone with (essentially) a background in experimental quantum optics. My great thanks also goes to both Tom{\' a}{\v s} Brauner and Anders Tranberg for their help in the papers I co-authored with them. I don't think I ever learned more than in the weeks I spend hosted by the two of you.

Thinking now there is a horde of other names here at the department to whom this thesis owes some thanks. From the bones of the latex code seen here, stacks of questions about Mathematica and the irrelevant questions one asks to begin thinking about the answer in a new way, to help getting through teaching duties and what all those forms actually wanted me to write, there was a lot to be done that I didn't know how to do. Thanks to all that helped.

And thank you to friends and family for allowing me to keep on going.

\cleardoublepage\addcontentsline{toc}{chapter}{\bf List of papers}
\chapter*{List of Papers}

  \textbf{Paper~\cite{wn1}}\\
  Jens O. Andersen, William R. Naylor and Anders Tranberg.\\
  \newblock {\em Chiral and deconfinement transitions in a magnetic background using the functional renormalization group with the Polyakov loop.}\\
  \newblock Journal of High Energy Physics, 1404:187, 2014.
  
  \textbf{Paper~\cite{wn2}}\\
  Jens O. Andersen, William R. Naylor and Anders Tranberg.\\
  \newblock {\em Inverse magnetic catalysis and regularization in the quark-meson model.}\\
  \newblock Journal of High Energy Physics, 1502:205, 2015.
  
  \textbf{Paper~\cite{wn3}}\\
  Jens O. Andersen, Tom{\'a}{\v s} Brauner and William R. Naylor.\\
  \newblock {\em Confronting effective models for deconfinement in dense quark matter with lattice data.}\\
  \newblock Submitted to Physical Review D, arXiv:1505.05925, 2015

\cleardoublepage\addcontentsline{toc}{chapter}{\bf Notation}
\chapter*{Notation}
\label{ch:notation}

\subsubsection{Units}

\begin{itemize}

\item Unless specifically stated we will work in natural units, where $\hbar=c=k_B=1$.\footnote{Using this elegant convention we have only a single unit, that of the electron-volt (eV). Thus, for example, dimensionally energies and masses are given in eV, distances by $\trm{eV}^{-1}$ and velocities are dimensionless.}

\end{itemize}

\subsubsection{Vectors}

\begin{itemize}

\item Three-vectors will be denoted by boldface and their components denoted by Latin indices, $\mbf{p}=(p_1,p_2,p_3)$.

\item Minkowskian four-vectors will be denoted by lower-case letters and their components denoted by Greek indices, $p=(p_0,\mbf{p})$, and we use the signature $(+,-,-,-)$.

\item We will use the Einstein summation convention, 
thus we have
\begin{align*}
	p_iq_i &= p_1q_1 + p_2q_2 + p_3q_3 \ep, \\
	p_\mu q_\mu &= p_0q_0 - p_iq_i \ep, \\
	P_\mu Q_\mu &= P_0Q_0 + p_iq_i \ep.
\end{align*}

\end{itemize}

\subsubsection{Definitions}

\begin{itemize}

\item The Pauli matrices in flavour/isospin (we will almost exclusively work with two flavors) space will be denoted by $\tau_i$ otherwise by $\sigma_i$ and, for reference, they are given by
\begin{equation*} \label{eq:pauli}
	\tau_1 =
  \begin{pmatrix}
  		0 & 1 \\
    1 & 0 \\
  \end{pmatrix} \ep,
  \quad
  \tau_2 =
  \begin{pmatrix}
    0 & -\im \\
    \im & 0 \\
  \end{pmatrix} \ep ,
  \quad
  \tau_3 =
  \begin{pmatrix}
    1 & 0 \\
    0 & -1 \\
  \end{pmatrix} \ep .
\end{equation*}
We also use $\sigma^\mu \equiv (1,\sigma_i)$ and $\bar{\sigma}^\mu \equiv (1,-\sigma_i)$.

\item The Gell-Mann matrices will be denoted by $\lambda_i$ and, again for reference, we give them here
\begin{align*} \label{eq:GellMann}
	\lambda_1 &=
  \begin{pmatrix}
  		0 & 1 & 0 \\
  		1 & 0 & 0 \\
  		0 & 0 & 0 \\
  \end{pmatrix} \ep,
  \quad
	\lambda_2 =
  \begin{pmatrix}
  		0 & -\im & 0 \\
  		\im & 0 & 0 \\
  		0 & 0 & 0 \\
  \end{pmatrix} \ep,
  \quad
	\lambda_3 =
  \begin{pmatrix}
  		1 & 0 & 0 \\
  		0 & -1 & 0 \\
  		0 & 0 & 0 \\
  \end{pmatrix} \ep, \notag\\
	\lambda_4 &=
  \begin{pmatrix}
  		0 & 0 & 1 \\
  		0 & 0 & 0 \\
  		1 & 0 & 0 \\
  \end{pmatrix} \ep,
  \quad
	\lambda_5 =
  \begin{pmatrix}
  		0 & 0 & -\im \\
  		0 & 0 & 0 \\
  		\im & 0 & 0 \\
  \end{pmatrix} \ep, \notag\\
  	\lambda_6 &=
  \begin{pmatrix}
  		0 & 0 & 0 \\
  		0 & 0 & 1 \\
  		0 & 1 & 0 \\
  \end{pmatrix} \ep,
  \quad
	\lambda_7 =
  \begin{pmatrix}
  		0 & 0 & 0 \\
  		0 & 0 & -\im \\
  		0 & \im & 0 \\
  \end{pmatrix} \ep,
  \quad
	\lambda_8 =
  \begin{pmatrix}
  		1 & 0 & 0 \\
  		0 & 1 & 0 \\
  		0 & 0 & -2 \\
  \end{pmatrix} \ep.
\end{align*}

\end{itemize}

\tableofcontents

\begingroup
\contentlayout

\chapter{Introduction}
\pagenumbering{arabic}

The work that comprises this thesis is mainly concerned with mapping out the location of the chiral and deconfinement phase transitions for strongly interacting matter. Confinement is perhaps the defining feature of quantum chromodynamics (QCD) and yet it remains poorly understood. This is mostly because of the difficulty in studying the non-perturbative regime of QCD. This thesis, and indeed the whole field of effective models for QCD, has arisen essentially as a response to this basic difficulty.

We use the two-flavor\footnote{We will only use the two lightest quark flavors, the up ($u$) and down ($d$) quarks. The charm, top and bottom quarks all have masses over a GeV and thus are beyond the energy scales we are exploring. The strange quark has a mass of 100 MeV, and thus a number of authors using effective models use three flavors, see for example~\cite{buballa05,ciminale08,fu08,fukushima08,schaefer10,mao10},
however we are more interested in the behaviour of the transitions, than their exact values. Also, in paper~\cite{wn3} we will compare directly to lattice articles~\cite{cotter13,boz13} using two flavors.} quark meson (QM) and Nambu-Jona-Lasinio (NJL) models augmented by the Polyakov loop to study thermal equilibrium QCD for temperatures up to around 500 MeV. We primarily employ these models such that we can use standard calculational tools to make meaningful calculations for this region of the phase diagram of QCD. These models contain many of the same symmetries as QCD, in particular those relating to chiral symmetry. They do not, however contain any gluon degrees of freedom. The Polyakov loop is introduced to (a) add an approximate order parameter for deconfinement and (b) bring back into the model some of the effects of confinement (the suppression of quark degrees of freedom in the confined phase). In addition we include a constant classical background magnetic field, $B$, and investigate the effects this has on the phase diagram.

A complimentary approach to solving the non-perturbative regime of QCD is simply discretising space-time and solving QCD numerically, such an approach is called lattice QCD.\footnote{For a general review see~\cite{petreczky12}, for results showing the behaviour of the chiral transition at physics quark masses see for example~\cite{endrodi11}, and for the work we compare directly with in paper~\cite{wn3} see~\cite{cotter13,boz13}.} From our point of view this field has an important role to play, due to the difficulty in obtaining and interpreting experimental results and observations. Lattice QCD thus acts as an independent check of the results found in model studies. An important example is the behaviour of the chiral transition as a function of external magnetic field, which we will return to shortly. Lattice QCD is hindered by the sign problem (see for example Sec.~2 of~\cite{deforcrand09}), which makes calculations at finite baryon chemical potential almost impossible. Thus models have an important role to play in the building of the phase diagram.

\section{Papers and problems}
\label{letmegohome}

We present three papers in Part~2. The first paper is in essence a very simple paper. We follow up the work of Andersen and Tranberg~\cite{andersen12} by extending the analysis of the QM model to include the Polyakov loop. Both works use a finite background magnetic field and chemical potential and go beyond mean field in using the functional renormalization group (RG), see Sec.~\ref{sec:RG}. We concern ourselves in particular with the effect that the Polyakov loop has on the phase diagram, and with the various versions of the potential for the gauge sector.

During the time I have been undertaking my doctorate (2011-2015), there has been debate in the literature about the disagreement between lattice and model studies on the effect of a magnetic field on the chiral phase transition, and in particular the transition temperature $T_c$. Early lattice studies~\cite{delia10,delia11} showed in agreement with model results that $T_c$ increased with increasing $B$. However these results used heavy quark masses, later studies using light quarks, with physical values for the pion masses~\cite{endrodi11,bali12a,bali12b,bruckmann13,bali13} showed the opposite behaviour (that $T_c$ decreases with increasing $B$). This has led to a torrent of articles~\cite{fraga13,ferreira14a,farias14,ferreira14b,ferrer15,ayala15,mueller15} investigating how models can be tweaked to reproduce these later lattice results. Motivated by the contrasting results of \cite{fraga13} and the RG extension of that work, we primarily investigate how various renormalization schemes effect the results of QM model calculations where the Yukawa coupling is allowed to vary with $B$.

In the third paper we look in greater depth at comparisons between lattice and model calculations. As we wish to investigate Polyakov loop coupled models, we must turn from physical three color QCD to QCD with two colors (2cQCD). The advantage of this is that lattice 2cQCD is free of the sign problem, and thus we are able to compare to lattice simulations at both zero and large chemical potential. In addition we use to the NJL model. Further discussion of the properties of 2cQCD is left to Sec.~2 of paper~\cite{wn3} and references therein. 



\section{Quantum chromodynamics}
\label{sec:QCD}

With the stage set, let us turn to the problem at hand: quantum chromodynamics. QCD is the correct theory of the strong interactions. The starting point for the theory is a Dirac Lagrangian for each of the six quark flavors. Interactions between the quarks are introduced by coupling them to a gauge field, exactly as in quantum electrodynamics (QED). The symmetry group of the gauge field is SU(3). This allows for colorless baryons, bound states of three quarks (such as the proton), and importantly this gauge group is non-abelian, which is central to the physics that this theory describes. The Lagrangian of QCD is, simply,
\begin{equation}
	\mathcal{L}_\text{QCD}=\bar\psi(i \gamma^\mu D_\mu-m_0)\psi-\frac{1}{4}G^a_{\mu\nu}G_a^{\mu\nu} \ep,
	\label{eq:QCDsimple}
\end{equation}
where $D_\mu$ is the covariant derivative coupling the fermions to the gauge bosons - the gluons - with $D_\mu=\partial_\mu - i g A_\mu^a T^a$ and $G^a_{\mu\nu}$ is the non-abelian field strength. This compact form reflects the structure we have just described, but hides a lot of the mathematical detail. In more complete notation Eq.~(\ref{eq:QCDsimple}) is given by
\begin{equation}
	\mathcal{L}_\text{QCD}=\bar\psi^\alpha_i(i \gamma^\mu\partial_\mu-m_i)\psi^\alpha_i+g A_\mu^a\,\bar\psi^\alpha_i\gamma^\mu t^a_{\alpha\beta}\psi^\beta_i-\frac{1}{4}G^a_{\mu\nu}G_a^{\mu\nu} \ep ,
		\label{eq:QCD}
\end{equation}
with
\begin{align}
	&i = u,d,c,s,b,t\textrm{ is the quark flavor,}\notag\\
	&\alpha = 1,2,3\textrm{ is the gauge/color index,}\notag\\
	&a = 1, \ldots 8\textrm{ runs over the gluons,}\notag\\
	&t^a_{\alpha\beta} = \frac{1}{2}\lambda^a_{\alpha\beta}\textrm{ with $\lambda^a$ the Gell-Mann matrices,}\notag\\
	&G^a_{\mu\nu} = \partial_\mu A^a_\nu - \partial_\nu A^a_\mu + g f^{abc}A^a_\mu A^b_\nu\textrm{ where $f^{abc}$ are the structure constants of SU(3).}\notag
\end{align}	

Given the definition of the structure constants as $[T^a,T^b]=\im f^{abc}T^c$ we see that the non-commutativity of the group immediately leads to an interaction term for the gluons.

The fact that the gauge bosons of QCD are self interacting leads to asymptotic freedom, and equally, color confinement.\footnote{This part is mostly based on Secs.~16.6 and 7 and 17.2 of Peskin and Schroeder~\cite{peskin95}, see there for greater detail. We should note that asymptotic freedom also requires the negativity of the beta function, and thus we must also have a sufficiently low number of fermion flavors (see Eq.~(\ref{eq:betaSUN})).} Loosely, as a color-neutral hadron - either a quark-antiquark pair (a meson) or three quarks of color red, blue, green (a baryon) - is pulled apart the color field holding them together increases in strength. The further apart the components of the hadron are, the stronger the force between them, and thus the greater the energy in the field. At some point (on the order of fermi) the energy in the field is so great that it is energetically favourable to create a quark-antiquark pair to create (more) localised color neutral objects.

Mathematically these concepts are most simply expressed through the running of the QCD coupling constant $\alpha_s=g^2/(4\pi)$. $g$ is the gauge coupling in Eq.~(\ref{eq:QCD}) and is related to the beta function (which describes how the coupling within a theory changes with energy scale $Q$) via
\begin{equation}
	\beta(g)=\frac{\pa g}{\pa \trm{log}(Q/\mu)} \ep,
	\label{eq:betaFunction}
\end{equation}
where $\mu$ is an arbitrary renormalization point. For an $SU(N_c)$ gauge theory with fermions in the fundamental representation the beta function is given by
\begin{equation}
	\beta(g)=-\frac{g^3}{(4\pi)^2}\left( \frac{11}{3}N_c-\frac{2}{3}N_f \right) + \mcal{O}(g^5) \ep .
	\label{eq:betaSUN}
\end{equation}
This result already tells us that QCD, which has $N_c=3$ and $N_f=6$ has a negative beta function and thus will be asymptotically free. Combining Eqs.~(\ref{eq:betaFunction}) and (\ref{eq:betaSUN}) with the definition of $\alpha_s$ we arrive at
\begin{equation}	
	\alpha_s(Q)=\frac{\alpha_s(\mu)}{1+(b_0 \alpha_s(\mu)/2\pi)\trm{log}(Q/\mu)} \ep ,
	\label{eq:alpha1}
\end{equation}
where we have defined $b_0\equiv(11N_c-2N_f)/3$. This may be simplified by defining the energy scale
\begin{equation}
	\Lambda_\trm{QCD}\equiv\mu\,\trm{exp}\left\{ \frac{-2\pi}{b_0 \alpha_s(\mu)} \right\} \ep ,
\end{equation}
which indicates the energy at which $\alpha_s$ becomes large. Eq.~(\ref{eq:alpha1}) is then given by
\begin{equation}
	\alpha_s(Q)=\frac{2\pi}{b_0 \, \trm{log}(Q/\Lambda_\trm{QCD})} \ep ,
	\label{eq:alpha2}
\end{equation}
again encoding the coupling's dependence on the energy scale. It should be noted that this is only true for sufficiently low $\alpha_s$. Experimentally this general picture is confirmed, as is shown in Fig.~\ref{fig:QCDcoupling}.

\begin{figure}[htbp]
	\centering
	\hspace{0in}
 	\includegraphics[width=3.5in]{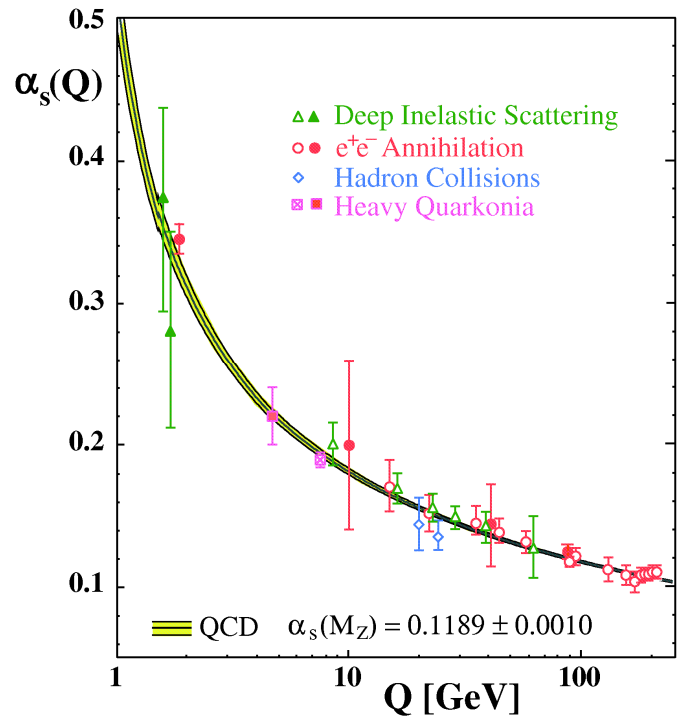}
	\caption[contentsCaption]{Summary of measurements of $\alpha_s(Q)$. The masses at which the charm and bottom quark freeze out is taken as 1.5 and 4.7 GeV respectively, at these points one can see small discontinuities in the theoretical prediction. Figure taken from Bethke, 2006~\cite{bethke06}.}t
	\label{fig:QCDcoupling}
\end{figure}

Despite this textbook knowledge of asymptotic freedom and the fine results shown in Fig.~\ref{fig:QCDcoupling}, an all encompassing definition of the deconfinement transition is still missing from QCD. It is clear that the transition is characterised by a change from hadronic degrees of freedom to quark degrees of freedom. We follow Fukushima~\cite{fukushima04} and others in coupling to the Polyakov loop (an order parameter for deconfinement in pure gauge theory~\cite{yaffe82a,yaffe82b}) which is introduced in  Sec.~\ref{sec:polyakov}.

Due to the very construction of QCD the theory clearly exhibits exact \tit{local} $SU(3)$ color symmetry. As a relativistic theory it is also invariant under the Poincar{\'e} group, that is both Lorentz transformations and translations in space and time. The Dirac spinor appears in combination with the Dirac adjoint, $\bar{\psi}=\psi^\dagger\gamma^0$, hence the Lagrangian is invariant under a $U(1)$ rotation of the quarks, resulting in the conservation of Baryon number. If we are lucky enough to have read Maggiore's fantastic introduction to quantum field theory (QFT) \cite{maggiore05} we will also recall that we built our Dirac spinor out of a pair of left handed and right handed Weyl spinors such that it would be invariant under a parity transformation, $(t,\mbf{x})\rightarrow(t,-\mbf{x})$, and of course this is carried over to the full theory, which is also time reversal and charge conjugation invariant. In the \tit{chiral representation} the Dirac spinor and the gamma matrices are simply given as
\begin{equation}
	\psi=
		\begin{pmatrix}
		\psi_L \\
		\psi_R \\
		\end{pmatrix} \ep,
	\qquad \qquad
	\gamma^\mu=
		\begin{pmatrix}
		0 & \sigma^\mu \\
		\bar{\sigma}^\mu & 0 \\
		\end{pmatrix} \ep ,
\end{equation}
where $\psi_{L(R)}$ is a left (right) handed Weyl spinor (and $\psi$ is a Dirac spinor). In this representation it is easy to see that if we neglect the mass term the quark sector is not only invariant under a total $U(1)$ rotation, but actually by two independent $U(1)$ rotations on the left and right handed parts:
\begin{equation}
	\psi_L\rightarrow e^{i \theta_L}\psi_L \ep ,
	\qquad \qquad
	\psi_R\rightarrow e^{i \theta_R}\psi_R \ep .
\end{equation}
We may choose to instead encode these two rotations as
\begin{equation}
	\psi \rightarrow e^{i \alpha}\psi \ep ,
	\qquad \qquad
	\psi \rightarrow e^{i \beta \gamma^5}\psi \ep ,
\end{equation}
where we have defined $\alpha = (\theta_L + \theta_R)/2$ and $\beta = (\theta_R - \theta_L)/2$, the associated symmetries are then denoted $U(1)_V$ (`vector' - this corresponds to baryon number) and $U(1)_A$ (`axial') respectively.
Using only the two lightest quark flavours we define
\begin{equation}
	q =
	\begin{pmatrix}
	u \\
	d \\
	\end{pmatrix}\ep ,
\end{equation}
where $u$ and $d$ are Dirac spinors for the respective quarks. Again neglecting the mass term the fermionic sector of Eq.~(\ref{eq:QCDsimple}) can be broken into two equations for the left and right handed parts with a global $U(2)_L\times U(2)_R$ symmetry. It is customary to break this down into $SU(2)_L\times SU(2)_R\times U(1)_V \times U(1)_A$.\footnote{The axial phase symmetry, $U(1)_A$, present in the classical theory is actually spontaneously broken in the QCD in a mechanism driven by instantons~\cite{thooft76a, thooft76b}. For our purposes we ignore, for simplicity, this caveat, however it is noted that in both the NJL and QM models one can simply augment the Lagrangians given by a term violating this symmetry.} This extension of the chiral symmetry across two massless flavors will be centrally important to us when we go to build our Nambu-Jona-Lasinio (NJL) and Quark-Meson (QM) models. Note that although this is hidden in our notation $q$ is also a three-vector in color space. 

At low temperatures this chiral symmetry is broken by the condensation of $q$-$\bar{q}$ pairs into the \tit{chiral condensate}, $\langle\bar{q}q\rangle$. The chiral condensate then serves as an order parameter for chiral symmetry. Thus, in the ground state the symmetry is reduced with $SU(2)_L\times SU(2)_R\rightarrow SU(2)_V$. At some sufficiently high temperature, $T_c$, the thermal energy will melt the chiral condensate and our chiral symmetry will be restored. This dynamic will be seen time and time again in the papers in Part~2 for various values of the chemical potential, the external magnetic field, the Yukawa coupling, the cutoff value, the parameter set for the gluonic sector and an external source term.

\section{Phase diagram of QCD}

The phase diagram of QCD is usually given in the temperature ($T$)-baryon chemical potential ($\mu_B$) plane. Figure~\ref{pd}, taken from~\cite{hands01}, gives a quantitative overview of this phase diagram.
\begin{figure}[htbp]
	\centering
	\hspace{0in}\hspace*{-0.4in}
 	\includegraphics[width=4.8in]{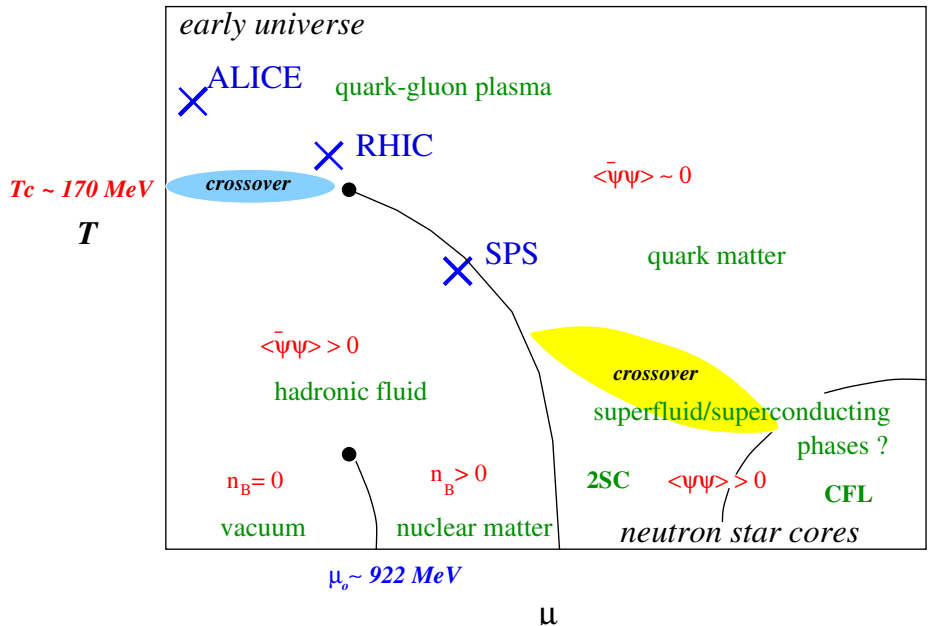}
	\caption[contentsCaption]{A qualitative representation of the phase diagram of QCD in the $T$-$\mu_B$ plane. Figure taken from~\cite{hands01}.}
	\label{pd}
\end{figure}
Figure~\ref{pd} also denotes the regions of particular experimental significance. Firstly, we have the heavy ion collisions at the ALICE detector and the Super Proton Synchrotron (SPS) (both at CERN), the Relativistic Heavy Ion Collider (RHIC) in Brookhaven. We refer the reader to Ref.~\cite{wong94} for an introduction to heavy ion collisions. Beyond this, the region of very high temperature was created in the very early universe, and neutron stars have the correct properties such that they may be used for studying the low temperature high density  region of the phase diagram. It is noted that observations of the quark-gluon plasma that presumably existed in the early universe are not directly observable, as it would have only existed long before cosmic microwave background was created and obscures our view of earlier times~\cite{hands01}.

The phase diagram may be split into three main regions, firstly at low $T$ and $\mu_B$ we have the hadronic phase, at sufficiently high temperatures the physics is dominated by a quark-gluon plasma and at high $\mu_B$ there exist several possible phases of quark matter. The hadronic phase is characterised by confinement of the quarks, and spontaneously broken chiral symmetry. The region is bounded by the `QCD transition', where both chiral symmetry is restored and matter becomes deconfined. Above this transition temperature we have a region dominated by the quark-gluon plasma, which is believed to be exactly what its name suggests. Indirect observation of this state was reported by CERN in 2000 and direct evidence from RHIC in 2005~\cite{brahms05,phenix05,back05,star05}. At high density there is believed to be found a color-flavor locked (CFL) phase, where quarks form Cooper pairs resulting in color-superconductivity. A number of other possible phases of quark matter have been proposed, see~\cite{alford08} for a review.

The extension of this phase diagram into the plane of magnetic fields was briefly discussed in Sec.~\ref{letmegohome} and we return to this in the next section. For further details see the recent review by Andersen et al.~\cite{andersen14}.



\section{Finite magnetic fields}

In papers~\cite{wn1} and \cite{wn2} we focus on QCD in a strong magnetic background. The phase diagram of QCD is experimentally accessible in the high $T$ ($\sim$250 MeV) and low baryon chemical potential through heavy ion collisions, and can be observed at high $\mu_B$ and low $T$ through observations of heavy stars, in particular neutron stars and magnetars. However in both heavy ion collisions and magnetars, very strong magnetic fields can arise. Additionally the early universe was subjected to high magnetic fields and thus it's understanding is also aided by such studies.

In heavy ion collisions (which are exactly what their name suggests) if the two nuclei do not collide exactly head on, but rather with some finite impact parameter, $b$ (see Fig.~\ref{fig:heavyIon}), then the geometry dictates that a magnetic field will be created. As the particles are ultrarelativistic (in the calculations presented below $\gamma\sim100$) the ions are Lorentz contracted such that one should imagine two (slightly offset, and electromagnetically charged) cr{\^e}pes travelling towards one another, as given by the large circles in Fig.~\ref{fig:heavyIon}. The motion of the charges is (to me) so like that of a curl operator ($\mbf{\nabla}\times$) it seems evident from Faradays law that we must have $\pa_t\mbf{B}\neq0$. The question for us is, is the magnitude of this induced (electromagnetic) magnetic field\footnote{In this thesis we will simply use `magnetic field' or similar, as we always refer to electromagnetic fields, and never chromomagnetic fields.} strong enough to have an effect upon the states of our strongly (chromodynamically) interacting system.

In the appendix of \cite{kharzeev08} the magnetic field strength is estimated based on the charge density of the nuclei and is shown in Fig.~\ref{fig:hia}. We see that the field's strength reaches a few $m_\pi^2$ ($\sim10^{18}$ Gauss $\sim10^{14}$ Tesla) with a centre of mass energy of 200 GeV per nucleon, which is approximately that used in RHIC. This result is in agreement with the calculations of~\cite{skokov09} based on the ultrarelativistic quantum molecular dynamics (UrQMD) model (see~\cite{bleicher99}) shown in Fig.~\ref{fig:hib}. Here the authors also show that $eB\sim10^{-1} m_\pi^2$ at the SPS (CERN) collider and $eB\sim15 m_\pi^2$ at the LHC. Clearly these fields are strong enough to impact the phase diagram of QCD.

The strength of magnetic field within magnetars is somewhat unclear. On the surface field strengths of $\sim10^{14}$ Gauss are observed~\cite{duncan92,kouveliotou99} and its possible that fields of the same order as that seen in heavy ion collisions are present in the cores of these stars.

\begin{figure}[htbp]
	\centering
	\hspace{0in}
 	\includegraphics[width=3.5in]{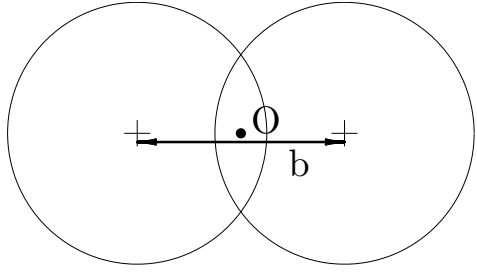}
	\caption[contentsCaption]{Basic geometry of heavy ion collisions. The circles represent the edges of the ions travelling into, and out of, the page. The impact parameter $b$ is denoted along with the centre $O$ of the collision, where magnetic field calculations are typically made.}
	\label{fig:heavyIon}
\end{figure}


\begin{figure}[htbp]
	\centering
	\subfigure[Note that $10^4$ MeV$^2$ corresponds to $\sim$0.5 m$_\pi^2$. Figure taken from Kharzeev et al.~\cite{kharzeev08}.]{
		\includegraphics[width=2.3in]{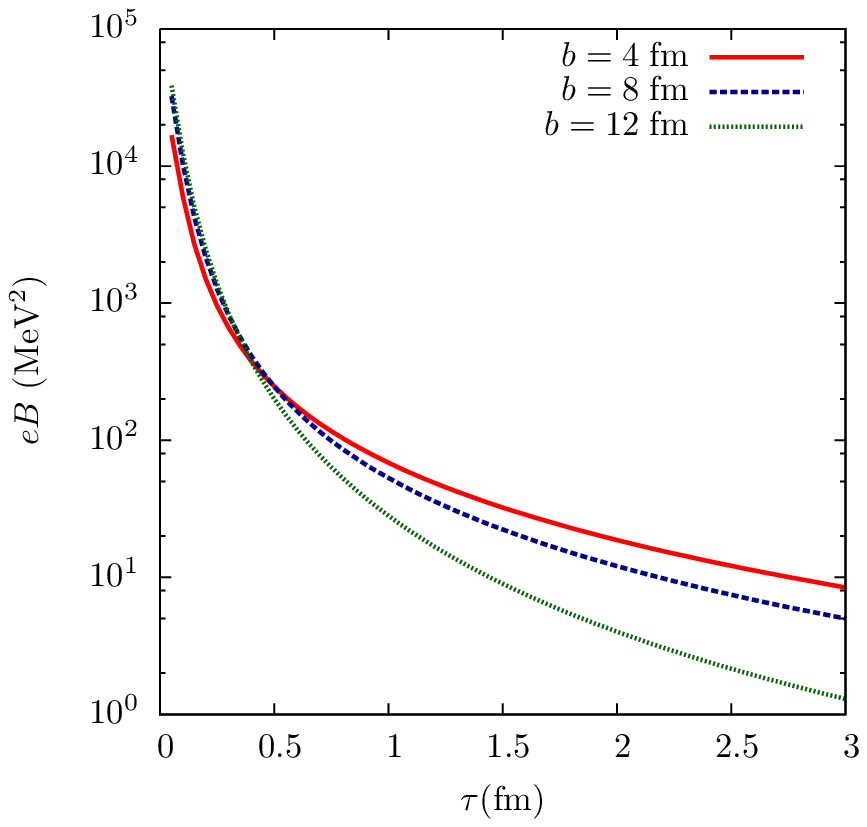}
		\label{fig:hia}}
	\subfigure[1 ev. is a single run of the UrQMD model, 100 ev. is the average of 100 runs. Figure taken from Skokov et al.~\cite{skokov09}.]{
  		\includegraphics[width=2.3in]{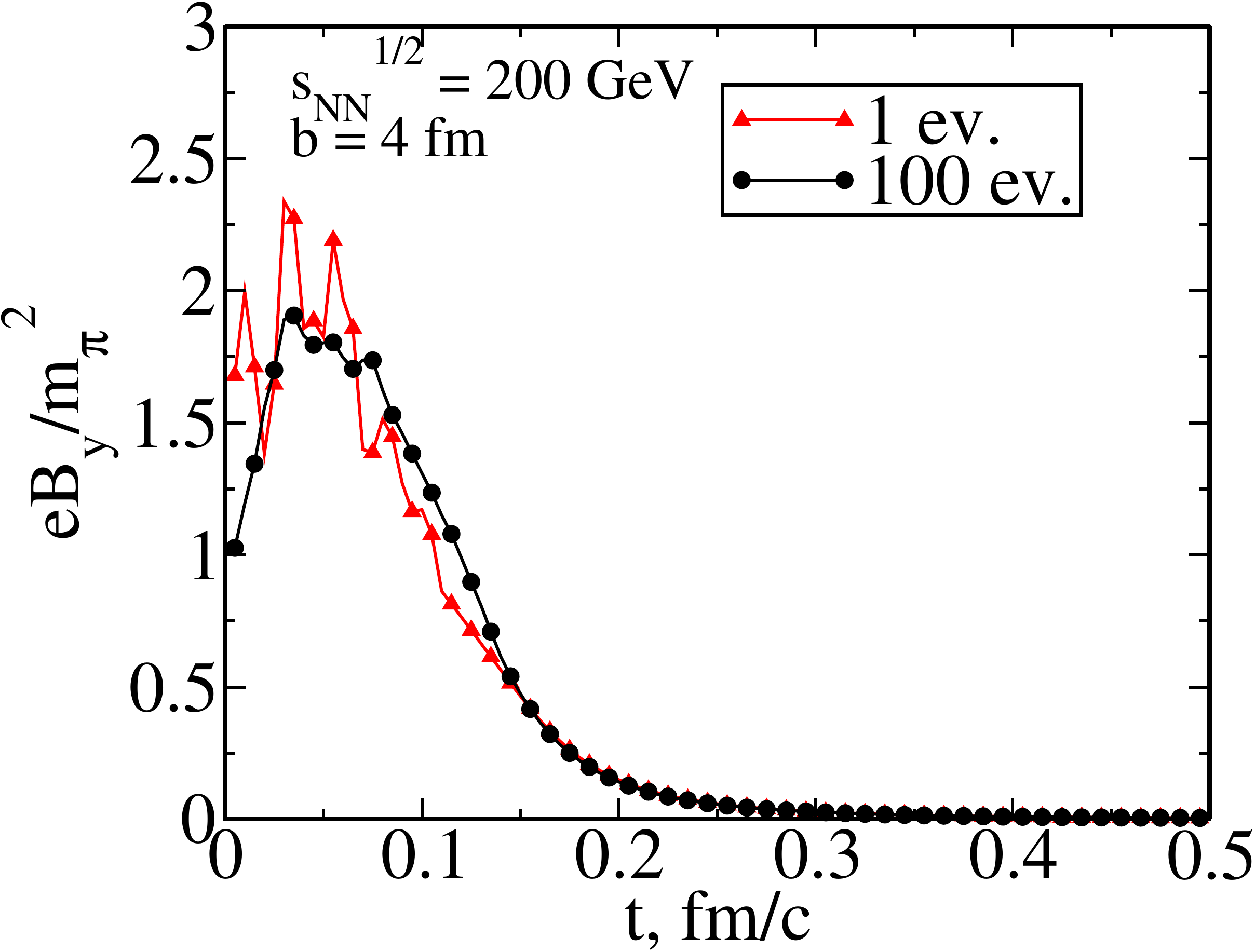}
  		\label{fig:hib}}
	\caption[caption for double]{Magnetic field strength at $O$ for a gold-gold (Z=79) collision with a centre of mass energy of 200 GeV from \cite{kharzeev08} and \cite{skokov09}. The impact factor(s) are given in the plots.}
	\label{fig:hi}
\end{figure}



\section{Thesis outline}

In Chapter~\ref{cha:mechanisms} we discuss the two models for QCD that we use, first the QM model and then, more briefly the NJL model. Here we will focus on how the models are constructed and which symmetries from QCD they contain. We also introduce the coupling of these models to the external background gauge (from which the Polyakov loop is defined) and magnetic fields.

In Chapter~\ref{cha:fundamentals} we give the the calculational methods used. Introducing first the basics of statistical thermal field theory, and defining the phase transitions we encouter time and time again in our papers. We introduce the functional renormalization group and give some calculational details. In this chapter we leave out a MF calculation of the QM model, as a very thorough explanation is given in~\cite{andersen11}.

Finally such that this thesis contributes more than simply fleshing out the theory around the papers presented in Part~2 we include a section on the numerical methods used in paper~\cite{wn1}. A great deal of time was spent during this thesis working through small numerical problems encountered in the calculations, and very little of this information is available in the literature. The hope of this section is that a prospective student can perform such calculations in the future with far greater ease and awareness of the possible potholes.


\chapter{Model approach to QCD}
\label{cha:mechanisms}

There is a small section on page 340 of Zee's QFT in a nutshell~\cite{zee10} titled `The Lagrangian as a mnemonic', which I have used a number of times to explain why it is effective theories are such a natural response to solving low temperature (non-perturbative) QCD. In the fifties many physicists felt that field theory was irrelevant for studying QCD and instead derived results general principles and in particular focusing on the symmetry structure of QCD. Then people realised they could take this approach further and encode those symmetries within an effective Lagrangian containing the symmetries they were studying. This comes with the obvious bonus that one can again use the well established machinery of QFT to analyse the Lagrangian.

Here we mostly aim to motivate and illustrate the construction of the Lagrangians of the QM and NJL models. The phase diagram predicted by the QM model is given in~\cite{wn1} and of course in the literature, notably (from our perspective) in~\cite{schaefer07,skokov10,stokic10,andersen11,herbst10,andersen12,kamikado13a,skokov12,herbst13,kamikado13b}. In the case of the (3-color) NJL model see for example~\cite{klevansky89,buballa05}, and for the 2-color NJL model see~\cite{kogut99,brauner09,andersen10,amador13}.

%
\section{The quark meson model}
\label{sec:QM}

To represent the chiral sector of low energy QCD an obvious starting point is a massless Dirac Lagrangian using the quark doublet $q=(u\,d)^T$ as given in Sec.~\ref{sec:QCD}. To add a mass term, $m\,\bar{q}q$, whilst maintaining our chiral $SU(2)_L\times SU(2)_R$ symmetry we couple all four of our bilinears\footnote{The four bilinears are the left- and right-handed bilinears of both the $u$ and $d$ quarks, this is particularly evident in the chiral representation.} equally to an $SO(4)$ multiplet of mesonic fields (which in anticipation we will denote $(\sigma,\bm{\pi})$) and use the fact that $SU(2)_L\times SU(2)_R$ is locally isomorphic to $SO(4)$ to maintain our chiral symmetry. The exact form of the coupling is ${\bar q}(\sigma + i \gamma^5 \bm{\tau} \cdot \bm{\pi})q$, which is called a Yukawa coupling. Our full Lagrangian will thus be
\begin{equation}
	\mcal{L}_\trm{QM} = {\bar q}\left( i \gamma^\mu\partial_\mu + g(\sigma + i \gamma^5 \bm{\tau} \cdot \bm{\pi}) \right) q + \mcal{L}(\sigma,\bm{\pi}) \ep ,
	\label{eq:QM}
\end{equation}
where $\bm{\tau}$ are the Pauli matrices acting in isospin space and $\mcal{L}(\sigma,\bm{\pi})$ must be $SO(4)$ symmetric. To give the quarks a spontaneously generated mass we simply employ the linear sigma model;
\begin{equation}
	\mcal{L}(\sigma,\bm{\pi}) = \frac{1}{2}\left( (\pa\sigma)^2 + (\pa\bm{\pi})^2 \right) + \frac{m^2}{2}\big( \sigma^2 + \bm{\pi}^2 \big) - \frac{\lambda}{4}\big( \sigma^2 + \bm{\pi}^2 \big)^2 - h\sigma \ep ,
\end{equation}
which has built in spontaneous symmetry breaking (assuming $m$ and $\lambda$ are positive). Thus in the vacuum we may choose $\langle \bm{\pi} \rangle = 0$ and $\langle \sigma \rangle > 0$. We have added the term $-h\sigma$, which we ignore for now, but will return to later. Additionally we have introduced four couplings, $g$, $m$, $h$ and $\lambda$, which we must fix to observables.\footnote{In addition we will later introduce a renormalization scale. The procedure of parameter fixing is well described in the literature and the papers in the second Part of this thesis. See~\cite{andersen11} for an example in a mean field calculation and~\cite{wn1} using the functional renormalization group.}

If we look at the coupling introduced between the $u$ and $d$ quarks via mesonic fields (and as the nomenclature already suggests) the 4-vector of mesons represents the sigma meson\footnote{The $\sigma$ meson is a very broad resonance, with the particle data group giving its mass in the range 400-550 MeV~\cite{olive14}. In~\cite{caprini06} the mass range is constricted to $400\pm6$ MeV.} and the three pions. In the case of the pions the physically relevant degrees of freedom are $\pi^0=\pi_3$ and $\pi^\pm=(\pi_1\pm i\pi_2)/\sqrt{2}$.

We may now simply follow the textbook examples of~\cite{zee10} (Chap. VI.4) or \cite{peskin95} (Chap. 11.1) for spontaneous symmetry breaking. Using the SO(4) symmetry we expand out fields around a non-zero minimum, $v=\sqrt{m^2/\lambda}$, of the mesonic potential in the $\sigma$ direction. We thus make our sigma field massive, and the pion fields become Goldstone bosons. Breaking this symmetry breaks the chiral symmetry of the model and thus our original $SU(2)_L\times SU(2)_R\times U(1)_V \times U(1)_A$ is broken down to $SU(2)_V\times U(1)_V \times U(1)_A$. Here the subscript $V$ stands for vector, but it could also be interpreted as the symmetry group for isospin. Note that when we expand our fields around $v$ we generate a quark mass term which is symmetric in the $u$ and $d$ flavors: $gv\,\bar{q}q$.

Currently our pions must be massless as they are Goldstone bosons, however in reality they have masses of approximate 140 MeV. Such that our model can reproduce these we have added the explicit chiral symmetry breaking term $-h\sigma$. In practice this gives us a parameter with we may tune the pion mass, however this must be treated with some care in the RG approach, as noted in Sec.~\ref{sec:RG}.

In paper~\cite{wn1} we plot the phase diagram given by the QM model at finite $B$ and $\mu_B$ and including the Polyakov loop using a numerical solution to the RG equations (see Sec.~\ref{sec:RG}). In paper~\cite{wn2} we again explore the phase diagram of the QM model using mean field methods. In this paper we mainly focus on the effects of varying the Yukawa coupling parameter, $g$.

\subsection{Chemical potential}
\label{sec:mu}

For any conserved charge within a system we may add a chemical potential. In Papers~\cite{wn1} and \cite{wn3} we add a baryon number potential, $\mu_B$, corresponding to the conserved baryon number, although for simplicity we tend to use the quark chemical potential (which we refer to simply as the chemical potential, $\mu$, with $3\mu=\mu_B$). The number density for quarks is  $\bar{\psi}\gamma^0\psi$, thus in both the QM and the NJL models we simply supplement the Lagrangian (Eqs.~(\ref{eq:QM}) and (\ref{eq:NJL}) respectively) by the term $\mu\,\bar{q}\gamma^0q$.


%
\section{The Nambu Jona-Lasinio model}
\label{sec:NJL}

A standard textbook introduction to QCD (for example~\cite{zee10}) starts with a Dirac Lagrangian and promotes the global $U(1)$ symmetry to a \tit{local} $U(1)$ gauge symmetry and in doing so couples to the gluons. In the NJL model we demote this global symmetry back to a local one, and integrate out the gauge fields.\footnote{This is not how the model was initially created. It was first introduce by Yoichiro Nambu and Giovanni Jona-Lasinio in 1961~\cite{nambu61a,nambu61b} as a model for nucleons (not quarks). Hence the model includes chiral symmetry, but completely ignores confinement.} When removing the gauge fields we replace single gluon exchange with a four fermion interaction term with a phenomenological (and dimensionful) coupling parameter, $G$, as indicated in Fig.~\ref{fig:NJL}.
\begin{figure}[htbp]
	\centering
	\hspace{0in}
 	\includegraphics[width=2.5in]{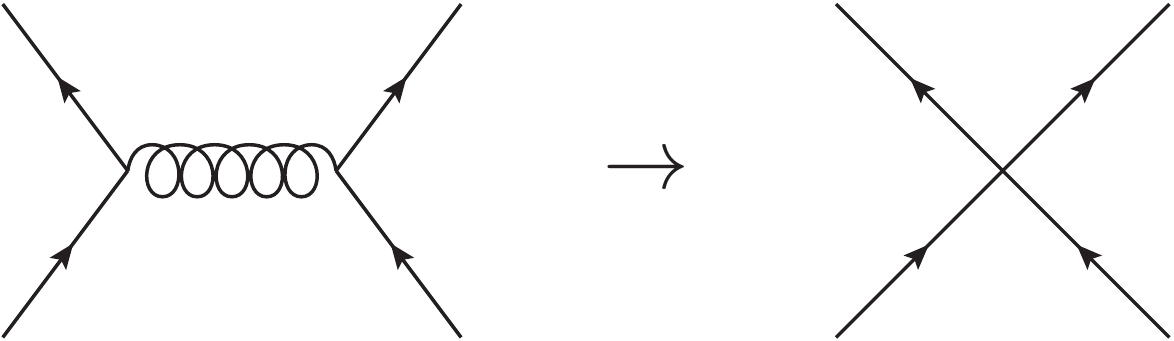}
	\caption[contentsCaption]{Integrating out the gluons from QCD the NJL model replaces single gluon exchange for a chirally symmetric four-fermion interaction.}
	\label{fig:NJL}
\end{figure}
To maintain chiral symmetry this interaction will be introduced as above, coupling all four bilinears equally, and thus the simplest chirally symmetric Lagrangian for a two-flavour NJL model (although adding a quark mass term $m_0$, which breaks chiral symmetry) is
\begin{equation}
	\mcal{L}_\trm{NJL} = {\bar q}\left( i \gamma^\mu\pa_\mu - m_0 \right) q + G\left[ (\bar{q}q)^2 + (\bar{q}i\gamma^5\bm{\tau}q)^2 \right] \ep .
	\label{eq:NJL}
\end{equation}
Often the interaction term is expanded to include addition scalar and pseudoscalar interactions. In addition one may arrange these terms so as to explicitly break the $U(1)_A$ symmetry that is found in Eq.~(\ref{eq:NJL}) (but not in the vacuum of QCD due to a quantum anomaly related to instantons~\cite{thooft76a,thooft86}). For an in depth article see~\cite{buballa05}, and see~\cite{andersen14} for an overview at finite $B$. In the case of 2cQCD paper~\cite{wn3} gives an overview of the possible operators and a expands on the standard NJL formalism, particularly in breaking the chiral symmetry of the interaction term with
\begin{equation}
	G\left[ (\bar{q}q)^2 + (\bar{q}i\gamma^5\bm{\tau}q)^2 \right] \rightarrow G (\bar{q}q)^2 + \lambda G (\bar{q}i\gamma^5\bm{\tau}q)^2 \ep ,
\end{equation}
where $\lambda$ is a new (dimensionless) parameter. Results from the NJL model at finite $B$ in 2cQCD are presented in~\cite{andersen13}.

\subsection{Mean field approximation}
\label{sec:meanfield}

In papers~\cite{wn2,wn3} we employ the mean field approximation. By this we mean the one-loop effective potential where we treat the mesonic fields at tree level, or equally, work in the large $N_c$ limit (but still use $N_c=3$ or $2$ as appropriate in our final expressions).

In the QM model the Lagrangian, Eq.~(\ref{eq:QM}), is already bilinear in the fermionic fields, so we may simply integrate over these. However the NJL model, as we have given it in Eq.~(\ref{eq:NJL}), contains terms with higher powers of the fermionic fields. Thus, to examine the low energy degrees of freedom we first make a Hubbard-Stratonovich transformation from the quark degrees of freedom to collective (mesonic) degrees of freedom. This is done by introducing a set of auxiliary fields, in the case of the simple Lagrangian above, have have only $\sigma$- and $\pi_i$-type degrees of freedom and thus we introduce the fields $\sigma=-2G\,\bar{q}q$ and $\pi_i=-2G\,\bar{q}i\gamma^5\tau_iq$. One then adds to the Lagrangian a term of the form $\delta\mcal{L}=-(\xi+2Gf(\bar{q},q))^2/4G$, where $\xi$ stands for the particular auxiliary field and $f(\bar{q},q)$ is chosen such that $\delta\mcal{L}=0$. In the example of $\sigma$ we have $\delta\mcal{L}=-(\sigma+2G\bar{q}q)^2/4G$. After making this transform the Lagrangian then takes the form
\begin{equation}
	\mcal{L}_\trm{NJL} = {\bar q}\left( i \gamma^\mu\pa_\mu - m_0 - \sigma - i\gamma^5\bm{\tau}\cdot\bm{\pi} \right) q - \frac{1}{4G}\left( \sigma^2 + \bm{\pi}^2 \right) \ep .
	\label{eq:NJL2}
\end{equation}
We are now able to exactly integrate out the (bilinear) fermionic fields and can deal simply with the mesonic fields (when working at mean field level).


%
\section{Finite B}
\label{sec:modelsB}

We raised two problems in the introduction that we wanted to address but are currently not included in our models, deconfinement and an external magnetic field. Here we introduce the basic mechanism for coupling to a magnetic field.

We examine the spectrum of a bosonic degree of freedom in a magnetic background. Considering only a classical constant background magnetic field with strength $B$ aligned along the $\hat{z}$ direction. In Sec.~2 of~\cite{andersen14} an equivalent derivation is given for fermions. The Lagrangian for our complex bosonic field is
\begin{equation}
	\mcal{L}_\trm{KG}=\frac{1}{2}(D_\mu\phi)^*D^\mu\phi - \frac{1}{2}m^2\phi^*\phi \ep ,
\end{equation}
where $D_\mu = \pa_\mu + ieA^\trm{EM}_\mu$. We have some gauge freedom in our choice of $A^\trm{EM}_\mu$, and choose $A^\trm{EM}_\mu=(0,0,Bx_1,0)$. The resulting equation of motion is
\begin{align}
	0 &= \left(D_\mu D^\mu + m^2 \right) \phi \\
	  &= \left(\pa_0^{\;2} - \pa_1^{\;2} - (\pa_2-ieBx_1)^2 - \pa_3^{\;2} + m^2 \right) \phi \ep,
\end{align}
and we will have a similar equation for the field $\phi^*$. Recalling the solution of the Klein Gordon equation, we suppose that $\phi$ takes the form $\phi=e^{i(p_0x_0-p_2x_2-p_3x_3)}f(x_1)$, which gives,
\begin{equation}
	  \left( -\pa_1^{\;2} + (\pa_2-eBx_1)^2 \right) \phi = \left(E^2 - p_3^{\;2} - m^2 \right) \phi \ep,
	  \label{eq:BHO}
\end{equation}
with $E=p_0$ and where we have been anticipating the physical meaning of the variables in our notation. Equation~(\ref{eq:BHO}) is that of a harmonic oscillator, and the solutions are the Hermite polynomials, although we will not be primarily concerned with the exact space-time dependence of the wave function. Of greater import is that the energy spectrum is
\begin{align}
	E_n^{\;2} &= m^2 + p_3^{\;2} + |e B|(2\ell + 1) & &\trm{(bosons)} \ep , \\
	E_n^{\;2} &= m^2 + p_3^{\;2} + |e B|(2\ell + 1 - s) & &\trm{(fermions)} \ep ,
\end{align}
where we have given the result for fermions in addition, with $s$ is the spin of the fermion and $\ell \in \mathbb{N}_0$ labels the energy of the associated Hermite polynomial. Note that we have gone (in spatial dimensions) from three continuous energy spectra, to a single one (along the $x_3$ axis) and a discrete variable, $\ell$. $\ell$ labels the Landau levels of the system, which denotes the energy associated with the `orbit' of the system. 

The result of this is that we must make the substitutions
\begin{align}
	\mbf{p}&\rightarrow p_z + (2\ell+1-s)|eB|\ep , & \int\!\frac{\trm{d}^3p}{(2\pi)^3} \rightarrow \frac{|eB|}{2\pi} \sum_{\ell=0}^\infty \int\!\frac{\trm{d}p_z}{2\pi} \ep,
\end{align}
(but with $s=0$ in the case of bosons) in our equations after transforming to momentum space. The RG implementation of this is given in~\cite{wn1}, and a detailed version of the calculation is given in App.~A of~\cite{andersen12}. Note that in the case the momentum substitution is done both in the derivatives of the effective action and in the regulator. The mean field implementation is given in~\cite{wn2}.


%
\section{Polyakov loop}
\label{sec:polyakov}

It is apparent that the models we are using do not contain gluons, and hence do not contain the mechanism for deconfinement as it stands in QCD. Additionally, at low temperatures the degrees of freedom in our two models include deconfined quarks, although we know that confinement means that the pions should dominate the physics. Thus following Fukushima~\cite{fukushima04} we add a constant temporal background gauge field, which as we will see suppresses the low temperature quarkyonic degrees of freedom. In addition the Polyakov loop, defined in terms of this background field, gives us an approximate order parameter for deconfinement.

The Polyakov loop, $\Phi$, is a traced Wilson loop around the periodic Euclidean time direction~\cite{polyakov78,suskind79}. The thermal Wilson line, $L$, is
\begin{equation}
	L(\mbf{x}) = \mcal{P}\trm{exp}\left[ i \int_0^\beta d\tau A_4(\mbf{x},\tau) \right] \ep,
\end{equation}
where $\mcal{P}$ is path ordering and $A_4=it_aA_0^a$. $\Phi$ (and its conjugate, $\bar{\Phi}$ \footnote{$L$ is in general a complex operator, thus we have two independent operators for the Polyakov loop, however both in 2cQCD and with $\mu_B=0$ $L\in\mathbb{R}$ and thus $\Phi=\bar{\Phi}$.}) is given in terms of $L$ as
\begin{align}
	\Phi &= \frac{1}{N_c} \langle \trm{Tr}_c L(\mbf{x}) \rangle \ep, & 	\bar{\Phi} &= \frac{1}{N_c} \langle \trm{Tr}_c L^\dagger(\mbf{x}) \rangle \ep,
\end{align}
where the trace is in color space. The free energy of a static quark is related to the expectation value of the Polyakov loop via $\langle\Phi\rangle=e^{-\beta F}$~\cite{holland00}. In QCD with static quarks adding a single test quark to the confined matter phase costs an infinite amount of energy, i.e.\ it is impossible and we have $\langle\Phi\rangle=0$. Thus the Polyakov loop forms an order parameter for deconfinement. It should be noted though, that this order parameter behaves in the opposite manner to most other order parameters in that it is zero in the low temperature (confined) state, and becomes non-zero at high temperature. In true QCD this is only approximately true, as a (dynamical) quark can couple to a (dynamical) antiquark and hence the energy cost is not infinite, and as such the Polyakov loop only serves as an approximate order parameter for deconfinement.

As stated the Polyakov loop is introduced by coupling to a constant temporal background gauge field. Thus we replace the derivative operator acting on the quarks as $\partial_\mu \rightarrow \partial_\mu - i\delta_{\mu0} t_a A_0^a$, where in the Polyakov gauge $t_a A_0^a$ can be written diagonally as $t_3A_0^3+t_8A_0^8$. The Fermi-Dirac distribution function is generalised to (see Sec.~VI of~\cite{andersen14})
\begin{align}
	n_F^+(\Phi, {\bar \Phi}; T, \mu) &= \frac{1+2{\bar \Phi}e^{\beta(E-\mu)} + \Phi e^{2\beta(E-\mu)}}{1+3{\bar\Phi} e^{\beta(E-\mu)}+3\Phi e^{2\beta(E-\mu)} +e^{3\beta(E-\mu)}}\ep, \label{n1}\\
n_F^-(\Phi, {\bar \Phi}; T, \mu)  &=n_F^+({\bar \Phi}, \Phi; T, -\mu) \ep , \label{n2}
\end{align}
which in the confined and deconfined limits becomes
\begin{align}
	n_F^+(\Phi, {\bar \Phi}; T, \mu) &= \frac{1}{1+e^{3\beta(E-\mu)}} & &(\Phi \trm{ \& } \bar{\Phi}\rightarrow 0)\ep,\\
	n_F^+(\Phi, {\bar \Phi}; T, \mu) &= \frac{1}{1+e^{\beta(E-\mu)}} & &(\Phi  \trm{ \& } \bar{\Phi}\rightarrow 1)\ep.
\end{align}
showing, as mentioned in~\cite{wn1} the suppression of the quarkyonic degrees of freedom in the confined phase.

When adding the Polyakov loop one must also add a potential for the pure gauge sector, which we do in all papers presented in Part~2. In paper~\cite{wn1} we give three such options and present results with all three potentials, as well as varying the parameter $\hat{\gamma}$, which controls the $\mu_B$ dependence of the transition temperature of the gauge potential. In paper~\cite{wn2} we investigate the dependence of the chiral transition on the parameter $T_0$. See the papers for the specific form of them implementation.


\chapter{Theoretical Formalism}
\label{cha:fundamentals}

In the papers that are presented below much of the formalism behind the calculations is left out. This tower of theoretical knowledge is built up from the foundations of mechanics, both quantum and classical, through field theory and thermodynamics and finally with the formalism of equilibrium thermal field theory. In this section we present for reference the relevant results from thermodynamics and thermal field theory (and an interesting phase transition found in the work for paper~\cite{wn1}). We also present the ideas behind the functional renormalization group and some interesting points that arise in the calculation of the flow equation of the effective potential. Finally we outline the renormalization schemes used, which are of particular import to paper~\cite{wn2}.

%
\section{Thermodynamic relations}
\label{sec:thermo}

The expectation values of the chiral and deconfinement order parameters, $\sigma$ and $\Phi$, can be found by minimising the grand canonical potential,
\begin{equation}
	\Phi_G = -T \trm{log}Z \ep,
	\label{camelbrains}
\end{equation}
with respect to these variables. $\Phi_G$ is an extensive quantity (it doubles with a doubling of volume) and the action is calculating over an infinite volume in space, thus we work with the grand potential's density:
\begin{equation}
	\Omega = \frac{\Phi_G}{V} = \frac{1}{\beta V} \trm{log}Z
	\label{eq:mariSoup}
\end{equation}
In addition to allowing the determination of the chiral transition temperature and the deconfinement transition temperature we may also define the pressure, $P$, entropy, $S$, particle number densities, $N_i$, and the energy density $\mcal{E}$ from $\Omega$ as
\begin{align}
	P &= -\Omega \ep , & S &= -\frac{\pa \Omega}{\pa T} \ep,\\
	N_i &= -\frac{\pa \Omega}{\pa \mu_i} \ep, & \mcal{E} &= \Omega + TS + \mu_i N_i \ep.
\end{align}
We have kept the particle number densities and chemical potentials general as we may have multiple chemical potentials. With two quark flavors with equal mass we may may include an isospin chemical potential, $\mu_I$, as well as a baryon chemical potential, but we set $\mu_I=0$ and only allow non-zero $\mu_B$.


%
\section{Phase transitions}
\label{sec:pt}

Studying the pressure or number density of a specific state of matter will in many cases be beyond our aims, instead we mostly focus on finding the location of the phase transitions. Usually this means finding the temperature at which chiral symmetry is restored, $T_c$, or matter becomes deconfined, $T_d$, for any given quark chemical potential or external magnetic field. The textbook method (which works more or less in the case of chiral symmetry) goes as follows. One identifies an appropriate order parameter, $O$, which will be zero in the symmetric phase, and non-zero in the phase with broken symmetry. One can then classify the phase transition by the way in which the order parameter goes to zero.

A phase transition is \tit{first order} where there is a discontinuous jump in $O$ from a finite value to zero. A \tit{second order} phase transition goes is an unbroken path from a finite value to zero but with a discontinuity in the first derivative of the order parameter. These `exact' phase transitions (where the order parameter goes to zero) occur when an exact symmetry of the Lagrangian is broken/restored. For example in Sec.~5 of paper~\cite{wn3} we see the breaking of $U(1)_V = U(1)_B$ with increasing $\mu_B$ and the eventual formation of a finite diquark condensate. In Fig.~\ref{fig:hi} we gives examples of the behaviour of the order parameter for various transition types. Figs.~\ref{fig:pta} and \ref{fig:ptb} give `exact' first and second order transitions respectively.

When the symmetry being restored is only approximate, as is the case of chiral symmetry in QCD (and our models) with non-zero quark masses, the phase transition also becomes approximate, with the order parameter transitioning to a low, but non-zero value. In these cases the critical temperature is often called the pseudo-critical temperature, although we will refrain from this excess. The definition of the critical temperature for an approximate first order phase transition is clear, it is the point at which we see a discontinuity in the order parameter, as is shown in Fig.~\ref{fig:ptc}. The definition of the critical temperature of a second order approximate phase transition is a little more ambiguous, as the the order parameter smoothly transitions from a high to low value. We will see in Sec.~\ref{sec:ptn} that this was numerically a difficult point to deal with. The generally accepted standard definition of the critical temperature is the inflection point, where $\pa^2 O/\pa T^2=0$, however it is often more convenient or precise to define the phase transition where the order parameter is at a half of it's zero temperature value. This is shown in Fig~\ref{fig:ptd}.

\begin{figure}[htpb]
	\centering
	\subfigure[Exact first order transition.]{
		\includegraphics[width=2.3in]{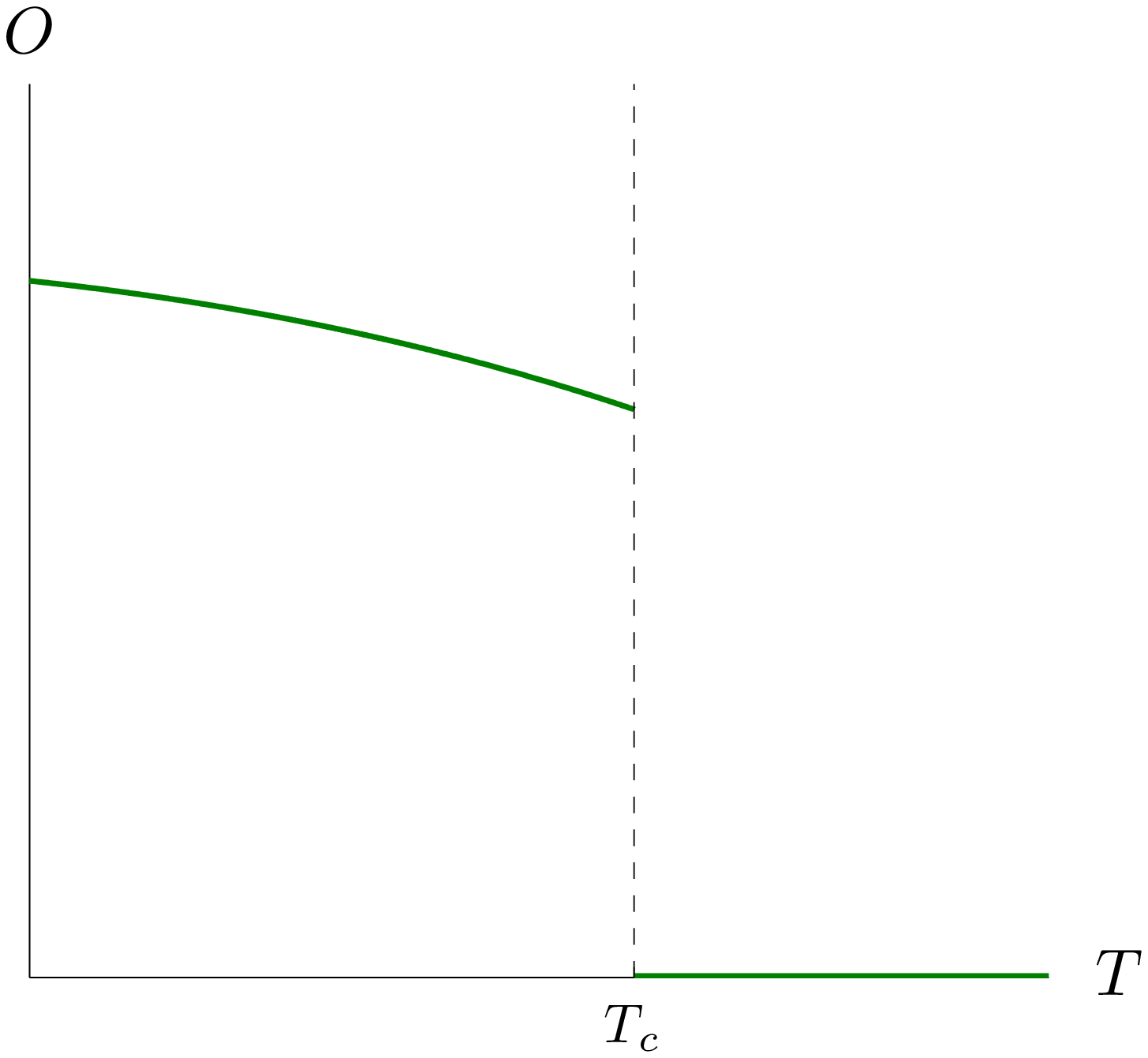}
		\label{fig:pta}}
	\subfigure[Exact second order transition.]{
		\includegraphics[width=2.3in]{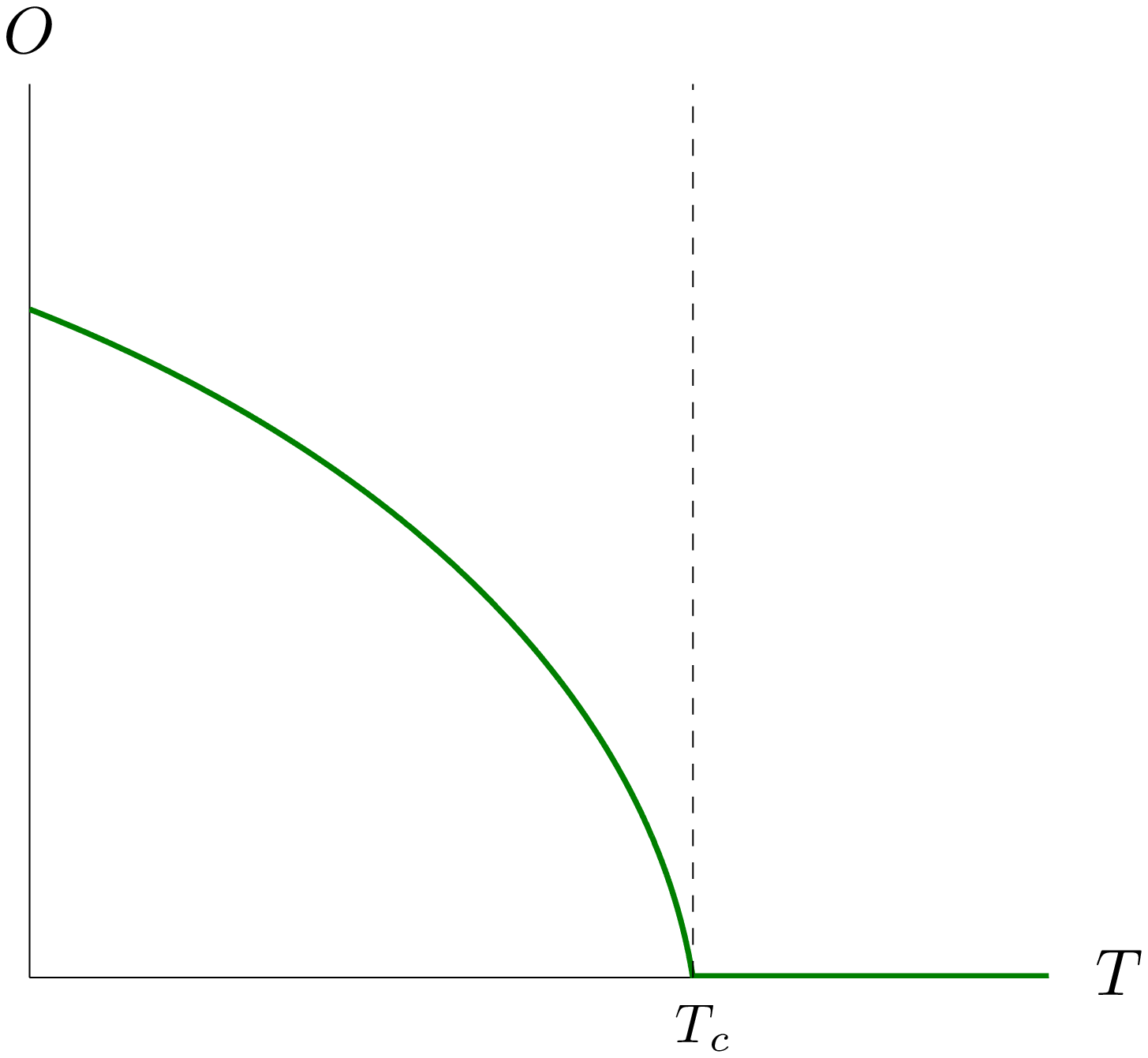}
		\label{fig:ptb}}
		\subfigure[First order transition restoring an approximate symmetry.]{
		\includegraphics[width=2.3in]{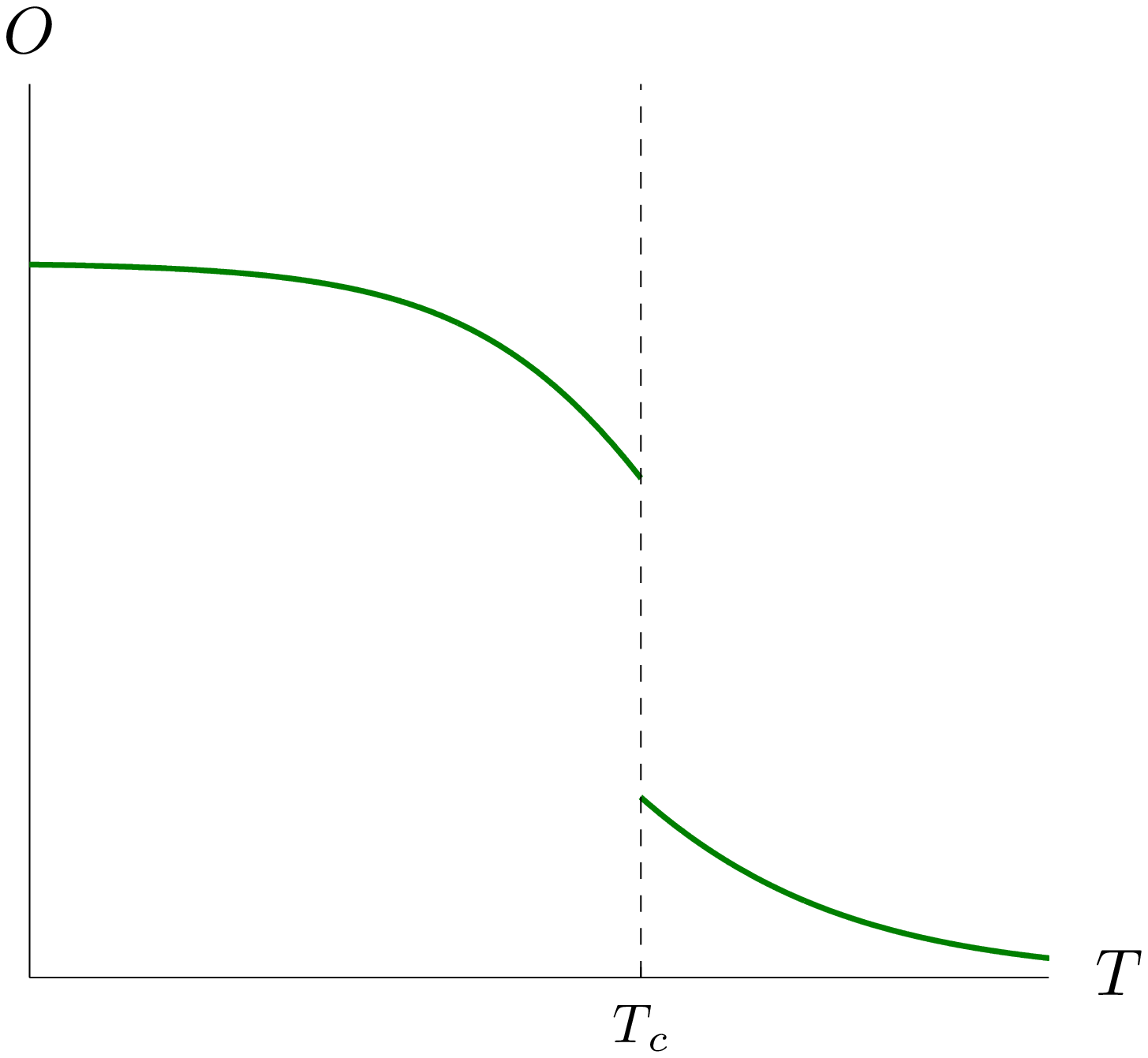}
		\label{fig:ptc}}
		\subfigure[Second order transition restoring an approximate symmetry.]{
		\includegraphics[width=2.3in]{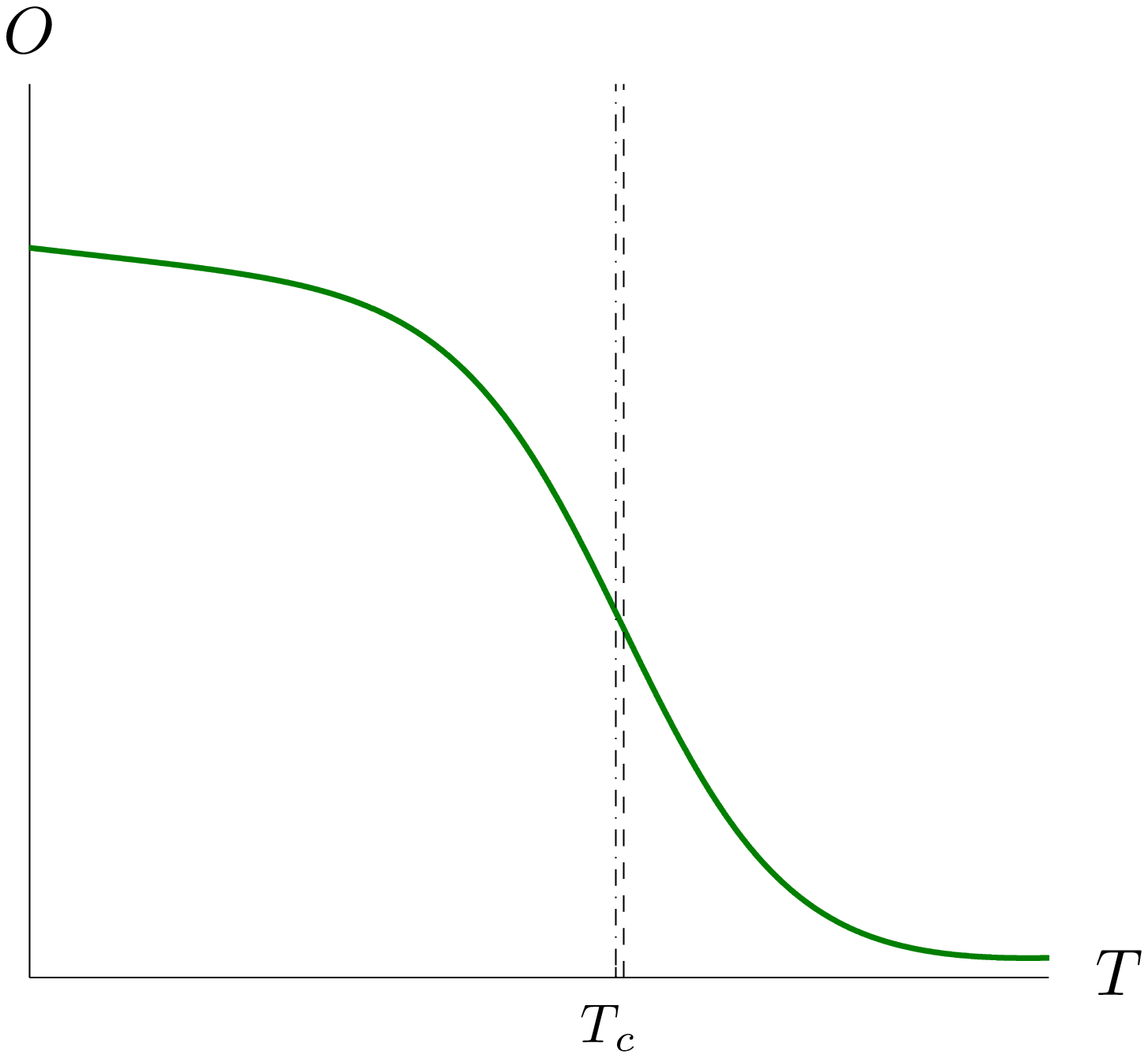}
		\label{fig:ptd}}
	\caption[caption for double]{Example curves for four different types of phase transitions. In (d) we show two transition temperatures, that at lower temperature (dashed-dotted) relates to the point at which $O(T)=O(0)/2$, while the higher temperature (dashed) line corresponds to the inflection point.}
	\label{fig:hi}
\end{figure}

Finally, as an interesting example originally observed in~\cite{schaefer05} (see specifically Figs.~1 and 2 of that paper), and also in our work towards~\cite{wn1}, there develops a bifurcation in the chiral phase transition at very low $T$ (below $\sim 20$ MeV). This occurs in the chirally symmetric quark meson model, evaluated with the functional renormalization group (see Sec.~\ref{sec:RG}), where the first order phase transition bifurcates to an approximate first order transition at slightly lower chemical potential and a second order transition at slightly higher $\mu_B$. The behaviour of the order parameter $\sigma$, the chiral order parameter, is shown in Fig.~\ref{fig:bi}. The reason for this, and if it is of any physical significance is unclear, although presumably it is completely washed out in QCD, where the quarks have non-zero bare mass.

\begin{figure}[htbp]
	\centering
	\hspace{0in}
 	\includegraphics[width=3.5in]{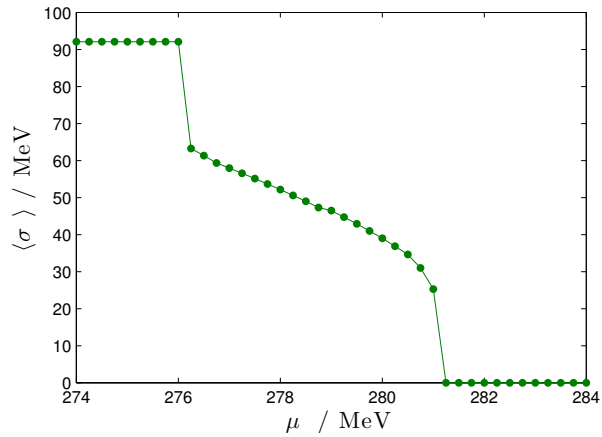}
	\caption[contentsCaption]{Chiral order parameter, $\langle \sigma \rangle$, at a function of $\mu$ at $T=10$ MeV. We see an approximate first order transition at $\mu\approx276$ MeV and a second order transition at $\mu\approx281$ MeV.}
	\label{fig:bi}
\end{figure}


%
\section{Thermal field theory}
\label{sec:thermal_field_theory}

We see from Eq.~(\ref{eq:mariSoup}) that the partition function is centrally important. Recalling our favourite book on thermal physics (for example~\cite{schroeder00}) we remember that the partition function is simply the sum over all Boltzmann factors, thus
\begin{equation}
	Z = \sum_\trm{all states}e^{-E(\trm{state})/k_BT} = \trm{Tr}\,e^{-\beta H}=\int\trm{d}\phi\bra{\phi}e^{-\beta H}\ket{\phi} \ep ,
\end{equation}
where $k_B$ is the Boltzmann constant, $H$ is the Hamiltonian and $\phi$ is a component in a complete set of continuous states. Using the essential result of path integral formulation of quantum field theory (QFT) (see for example~\cite{zee10}),
\begin{equation}
	\bra{\phi_F}e^{-i t H}\ket{\phi_I} = \int_{\phi_I,0}^{\phi_F,t}\mcal{D}\phi \;e^{i S[\phi]} \ep ,
\end{equation}
we see the partition function is a sum over closed ($x_I=x_F=x$) path integrals. In the formalism of thermal field theory this is made exact, and as a first step we rotate our time dimension to imaginary time, $t\rightarrow -i\tau$ and integrate from $0$ to $\beta$. When all of the $i$'s have been cancelled, and interpreting $\phi$ as a scalar field, we are left with
\begin{equation}
	Z=\int_{\phi(\mbf{x},0)=\phi(\mbf{x},\beta)}\mcal{D}\phi \; e^{-S_E[\phi]}\ep,
	\label{eq:asdf}
\end{equation}
where we integrate over all fields that satisfy $\phi(\mbf{x},0)=\phi(\mbf{x},\beta)$, that is all fields with period $\beta/n$ (with $n \in \mathbb{N}_+$) in imaginary time. Due to the anticommutivity of fermionic fields (see Sec.~2.5 of~\cite{kapusta06}) the fermionic version of Eq.~(\ref{eq:asdf}) is
\begin{equation}
	Z=\int_{\psi(\mbf{x},0)=-\psi(\mbf{x},\beta)}\mcal{D}\bar{\psi}\mcal{D}\psi \; e^{-S_E[\bar{\psi},\psi]}\ep,
	\label{eq:asdf2}
\end{equation}
with the sum now over anti-periodic fields. In both Eqs.~(\ref{eq:asdf}) and (\ref{eq:asdf2})
\begin{equation}
	S_E = \int_0^\beta \trm{d}\tau \int \trm{d}^3x \;\mcal{L}_E \ep,
\end{equation}
with
\begin{equation}
	\mcal{L}_E = -\mcal{L}(t\rightarrow-i\tau) \ep.
\end{equation}

In real space we see that we have an temporal integral over periodic boundary conditions. Once we have Fourier transformed to momentum space this will result in a sum over an infinite set of discretely spaced frequencies, the Matsubara frequencies, $\omega_n$ for bosons we have $\omega_n=2n\pi T$ and for fermions $\omega_n=(2n+1)\pi T$.

We now turn to the evaluation of the grand potential for free fermions. This serves both as demonstrational for the most essential analytic tools we have at our disposal: Fourier transforms, Matsubara sums and Gaussian integration. This will also elucidate the basis of the mean field approximation calculations given in papers \cite{wn2} and \cite{wn3}. In these papers we end up with a Lagrangian bilinear in the fermionic fields and with a number of higher order terms for the mesonic/bosonic fields. As stated in Sec.~\ref{sec:meanfield}, the mean field approximation the fluctuations of these bosonic fields are ignored and hence the results parallel the simplest case we present here.

We begin by Fourier transforming to momentum space using the relation
\begin{equation}
	\psi(\tau,\mbf{x}) = \frac{1}{\sqrt{\beta V}} \sum_n \int\!\frac{\trm{d}^3p}{(2\pi)^3} \psi_n(\mbf{p}) e^{i(\omega_n\tau+\mbf{p}\cdot\mbf{x})} \ep,
\end{equation}
thus the action is
\begin{equation}
	S_E = \sumint_\mbf{p} \bar{\psi}_n(\mbf{p}) \; G^{-1}_{n,\mbf{p}} \; \psi_n(\mbf{p}) \ep .
	\label{hozier}
\end{equation}
where we have defined both
\begin{align}
	\sumint_\mbf{p} &\equiv \sum_{n=\infty}^\infty \int\!\frac{\trm{d}^3p}{(2\pi)^3} \ep, \\
	G^{-1}_{n,\mbf{p}} &\equiv i \gamma^0\omega_n + \bm{\gamma}\cdot\mbf{p} + m \ep.
\end{align}
Substituting Eq.~(\ref{hozier}) into Eq.~(\ref{eq:asdf2}) we find for partition function, 
\begin{equation}
		Z=\int\mcal{D}\bar{\psi}\mcal{D}\psi \; \trm{exp}\left\{-\sumint_\mbf{p} \bar{\psi}_n(\mbf{p}) \; G^{-1}_{n,\mbf{p}} \; \psi_n(\mbf{p}) \right\} \ep.
\end{equation}
This is simply a Gaussian integral and has solution $Z=\trm{exp}\{\trm{Tr }\trm{ln }G^{-1}_{n,\mbf{p}}\}$. Utilising relation~\ref{eq:mariSoup} for $\Omega$ and that $\trm{Tr }\trm{ln }M = \trm{ln }\trm{det }M$ we have
\begin{align}
	\Omega &= -T \, \sumint_\mbf{p} \trm{ln }\trm{det } G^{-1}_{n,\mbf{p}} \notag\\
		&= -2T \,\sumint_\mbf{p} \trm{ln }\big[\omega_n^2 + \mbf{p}^2 + m^2\big] \notag\\
		&= -4 \int\!\frac{\trm{d}^3p}{(2\pi)^3} \left( \frac{1}{2}E_\mbf{p} + T\,\trm{ln}\Big(1 + e^{-\beta E_\mbf{p}}\Big) \right)  \ep,\label{jump}
\end{align}
where in the first step we have simply evaluated the determinant in Dirac space and in the final line we have defined $E_\mbf{p}\equiv\mbf{p}^2 + m^2$ and explicitly evaluated the Matsubara sum as follows. We often obtain a sum of the form
\begin{equation}
	X = \frac{1}{2}T \sum_{n=\infty}^\infty \trm{ln}(\omega_n^2 + a^2)
\end{equation}
which can be evaluated using complex contour integration, see Sec.~2.3.3 of~\cite{kyllingstad07}, or~\cite{leballac96}, to give
\begin{align}
	X &= \frac{1}{2}a + T\, \trm{ln}\big[1 - e^{-\beta a}\big] & &\trm{(bosons, $\omega_n=2n\pi T$)} \ep,\notag\\
	X &= \frac{1}{2}a + T\, \trm{ln}\big[1 + e^{-\beta a}\big] & &\trm{(fermions, $\omega_n=(2n+1)\pi T$)} \ep.\notag
\end{align}


%
\section{Functional renormalization group}
\label{sec:RG}

I find the ideas behind renormalization group flows particularly attractive. The Lagrangian details the physics present at the smallest length scale of the theory we are examining. The problem is reading off the physics over large length scales, where interactions may change the relevant degrees of freedom or symmetries and where the detailed properties of the system are not straightforwardly given.

We begin with the flow equation introduced by Wetterich~\cite{wetterich93} (for more general and pedagogical articles see~\cite{rosten12,berges00}) for the effective potential:
\begin{equation}
	\partial_k \Gamma_k[\phi,\psi] = \frac{1}{2}\trm{Tr} \left[ \frac{\partial_k R_{kB}}{\Gamma_k^{(2,0)} + R_{kB}} \right] - \trm{Tr} \left[ \frac{\partial_k R_{kF}}{\Gamma_k^{(0.2)} + R_{kF}} \right] \ep .
	\label{eq:herooo}
\end{equation}
$\Gamma_k^{(a,b)}[\phi,\psi]$ represents the $a$th functional derivative with respect to the mesonic fields and the $b$th with respect to the fermionic fields and $k$ is the current energy scale in the RG flow. $R_{kB(F)}$ is the bosonic (fermionic) regulator function, which we will introduce shortly. The basic idea of the equation is as follows: At some suitably high energy scale, $k=\Lambda_\trm{RG}$, the physics is simply that given by the classical action, i.e.\ the tree level potential\footnote{One has to be slightly careful in calculating absolute quantities, for example pressures, to include additive effects coming from the energies above $\Lambda_\trm{RG}$. In Paper \cite{wn1} we investigate the expectation values of the condensates, hence avoiding these problems.} (up to a factor $V\beta$). If we then imagine zooming out (lowering the energy scale) one `step' and re-writing our action in terms of new effective masses and interactions. Then zooming out again, and repeating the process, and so on. Imagining now these steps as infinitesimal we have a `flow' of the effective potential, with the high energy boundary condition being the tree level potential. When we reach $k=0$ then the effective potential will be equal to the full quantum action. Equation~(\ref{eq:herooo}) encodes (exactly) this flow from the classical to the quantum action.

The regulator functions, $R_{kB(F)}(p)$, control the flow equation. For the bosonic term in Eq.~(\ref{eq:herooo}) we use the form\footnote{For fermions we use
\begin{equation}
	R_{kF}(p) = \left( \sqrt{\frac{(p_0+i\mu)^2 + k^2}{(p_0+i\mu)^2 + \mbf{p}^2}} - 1 \right)(\gamma^\mu p_\mu + i\mu\gamma^0) \,\theta(k^2 - \tbf{p}^2) \ep ,
\end{equation}
which incorporates the same properties as the bosonic regulator, whilst also simplifying the calculation of the flow equation.}
\begin{equation}
	R_{kB}(p) = (k^2 - \tbf{p}^2) \,\theta(k^2 - \tbf{p}^2) \ep ,
\end{equation}
where $\theta(x)$ is the Heaviside step function. The cutoff ensures that modes far below the energy scale of the cutoff are heavy and decouple, whilst modes below but close to the cutoff are included in the flow. For modes above the cutoff ($\mbf{p}^2>k^2$), using this particular form of the regulator, the numerator of Eq.~(\ref{eq:herooo}) ensures that those modes are not included.  

In practice one cannot find an exact form of the flow equation for the QM model and approximations must be made. In paper~\cite{wn1} we use a truncated derivative expansion for the effective action in the local-potential approximation (where all of the terms in the expansion are left independent of the energy scale other than the effective potential for the mesonic fields and the Yukawa coupling). We then further approximate by ignoring the running of the Yukawa coupling. In addition we ignore any running of the Polyakov loop, and treat this as an independent classical background field as in the mean field case. We thus start with the following effective potential for the $U(4)$ symmetric mesonic field $\rho=\phi^2/2=(\sigma^2+\bm{\pi}^2)/2$
\begin{align}
	\Gamma_k[\rho] &= \int_0^\beta \trm{d}\tau \int \trm{d}^3x \; \bigg[ \frac{1}{2}\left( (\partial_\mu\sigma)^2 + (\partial_\mu\bm{\pi})^2 \right) + U_k(\rho) \\
	&\qquad\qquad\qquad + \bar{q} \gamma_\mu D^\mu q + g \bar{q}(\sigma + i \gamma_5 \bm{\tau}\cdot\bm{\pi})q\bigg],
\end{align}
where $D^\mu$ is the covariant derivative which couples to both the magnetic and gauge background fields as given in Secs.~\ref{sec:modelsB} and \ref{sec:polyakov}. As the effective potential we calculate is explicitly $U(4)$ symmetric the explicit symmetry breaking term ($h\sigma$) used to give the pions mass is dropped from the calculation initially (this amounts to not including it in the tree level potential), and instead adding it as a condition on the minimisation of effective potential when finding $\langle\sigma\rangle$ (see Chap.~\ref{cha:numerics} for details). The derivation of the flow equation follows that of App.~A of~\cite{andersen12}, other than the replacement of the Fermi-Dirac distribution function to their Polyakov loop extensions given in Eqs.~(\ref{n1}) and (\ref{n2}).\footnote{Note also in the derivation in Eq.~A.6 the second term is missing a factor $-2$, in Eq.~A.9 the term `$+U_k''$' should read `$+2\rho U_k''$' and Eq.~A.14 is missing a plus sign in the middle of the second line.}

Finally we must add the gluonic potential, but as stated this not treated as scale dependent and thus may simply be added to the final result for $U_{k=0}(\rho)$ before calculation the expectation values of $\sigma$, $\Phi$ and $\bar{\Phi}$.


%
\section{Renormalization}
\label{sec:reorm}

In paper~\cite{wn2} we compare the results of the QM model using dimensional regularization (commonly used in mean field model calculations) and a sharp cutoff (which we used in the RG calculations in paper~\cite{wn1}). Here we supplement the treatment in that paper of these two methods.

In Eq.~(\ref{jump}) we have the divergent integral of the form $\int\!\trm{d}^3p \, \sqrt{\tbf{p}^2+m^2},$\footnote{Recall that we arrived at this form from a single free fermion with mass parameter $m$. We are also ignoring a factor coming from a sum over the spin states, and are in a medium with $T=B=0$.} which remains in essence in the calculations in Part~2. A straightforward method for regulating the integral is to simply add a UV three-momentum cutoff, $\Lambda_\trm{UV}$, integrating from zero momentum up to the energy scale $\Lambda_\trm{UV}$ and ignoring anything above that. In this case the above integral,
\begin{equation}
	\int\!\frac{\trm{d}^3p}{(2\pi)^3} \sqrt{\tbf{p}^2+m^2} = \frac{1}{2\pi^2}\int_0^\infty\!\trm{d}p\; p^2\sqrt{p^2+m^2} \ep,
\end{equation}
becomes
\begin{align}
	\frac{1}{2\pi^2}\int_0^{\Lambda_\trm{UV}}\!\trm{d}p\; p^2\sqrt{p^2+m^2} &= \frac{1}{16\pi^2} \Bigg\{ \Lambda_\trm{UV}\sqrt{\Lambda_\trm{UV}^2+m^2}\,(2\Lambda_\trm{UV}^2+m^2) \notag\\
		&\qquad-m^4\,\trm{ln}\left[ \frac{\Lambda_\trm{UV} + \sqrt{\Lambda_\trm{UV}^2+m^2}}{m} \right] \Bigg\} \ep.
\end{align}
This somewhat crude scheme is far more useful than it might seem, as if all of the relevant physics occurs below this scale this will approximately amount to subtracting a (infinite) constant from the potential. However we must be clear that in doing this we are quite explicitly redefining our theory with $\Lambda_\trm{UV}$ as a parameter in the new theory, which other quantities (masses, couplings) will depend upon, although hopefully only weakly.

In dimensional regularisation (\cite{thooft73,weinberg73} or see~\cite{lepage05} for a more pedagogical article) we instead allow dimension of integration to vary, and then use this to find the form of the renormalization terms required to render the integral finite. We use the modified minimal subtraction scheme ($\overline{\trm{MS}}$) where,
\begin{equation}
	\int\!\frac{\trm{d}^3p}{(2\pi)^3} \rightarrow \left( \frac{e^{\gamma_E}\Lambda_\trm{DR}^2}{4\pi} \right)^\epsilon \int\!\frac{\trm{d}^dp}{(2\pi)^d} \ep,
\end{equation}
with $d=3-2\epsilon$ and $e^{\gamma_E}/4\pi$ is added to simplify later calculations. $\Lambda_\trm{DR}$ is the renormalization scale associated with the scheme, which is added to keep the integral dimensionally consistent while $d$ is varied. Switching to polar coordinate and then integrating and expanding in powers of $\epsilon$ up to zeroth order we obtain (see~\cite{andersen11} or pages 249-251 of~\cite{zee10})
\begin{align}
	&\left( \frac{e^{\gamma_E}\Lambda_\trm{DR}^2}{4\pi} \right)^\epsilon \int\!\frac{\trm{d}^dp}{(2\pi)^d} \sqrt{p^2+m^2} \notag\\
	&\qquad\qquad= \left( \frac{e^{\gamma_E}\Lambda_\trm{DR}^2}{4\pi} \right)^\epsilon \frac{2\pi^{d/2}}{\Gamma(d/2)}\frac{1}{(2\pi)^d}\int_0^\infty\!\trm{d}p \; p^{d-1} \sqrt{p^2+m^2} \notag\\
		&\qquad\qquad=-\frac{m^4}{32\pi^2}\left(\frac{\Lambda_\trm{DR}^2}{m^2}\right)^\epsilon \left(\frac{1}{\epsilon} + \frac{3}{2}\right) \ep.
\end{align}
Taking the limit as $\epsilon\rightarrow0$ we are left with a divergent term, $-m^4/32\pi^2\epsilon$ which may be removed by renormalizing the mass term. The final result is $-m^4/32\pi^2[\trm{log}(\Lambda_\trm{DR}^2/m^2)+(3/2)]$.

Here we have only considered a fermion in the vacuum. Although the finite temperature terms are inherently finite, the inclusion of a magnetic field leads to some additional divergencies. A divergent term explicitly dependent upon the magnetic field can be removed by renormalization of the magnetic field. In addition the sums over Landau levels are divergent and are regulated using the Hurwitz zeta function,
\begin{equation}
	\zeta(s,a)\equiv \sum_{k=0}^\infty\frac{1}{(k+a)^s} \ep.
\end{equation}
The sums are simply expressed in terms of $\zeta$ and then the result is from the analytic continuation of $\zeta$ to negative $s$ is inherently finite. For full details of the calculation see~\cite{andersen11}.

Although both schemes are equally applicable to the QM model, authors almost exclusively use dimensional regularization (in mean field treatments), presumably due to the weaker renormalization scale dependence of dimensional regularization and that the method preserves gauge invariance.


\chapter{Numerical methods}
\label{cha:numerics}

There is scant information in the literature on the numerical methods behind the calculation of the flow equation presented in~\cite{wn1}, something we are also guilty of. The task at hand is to evaluate the following equation:
\begin{align}
\partial_k U_k(\rho,&\Phi,\bar{\Phi};\,T,\mu,B) = {k^4\over12\pi^2}
\left\{
{1\over\omega_{1,k}}\left[1+2n_B(\omega_{1,k})\right]
+{1\over\omega_{k,2}}\left[1+2n_B(\omega_{2,k})\right]
\right\}\notag \\
&\nonumber
+k{|qB|\over2\pi^2}\sum_{\ell=0}^{\infty}{1\over\omega_{1,k}}
\sqrt{k^2-p^2_{\perp}(q,\ell,0)}\,\theta\left(k^2-p^2_{\perp}(q,\ell,0)\right)
\left[1+2n_B(\omega_{1,k})\right] \\
& \nonumber
-{N_c\over2\pi^2}k\sum_{s,f,\ell=0}^{\infty}
{|q_fB|\over\omega_{q,k}}
\sqrt{k^2-p^2_{\perp}(q_f,\ell,s)}\,\theta\left(k^2-p^2_{\perp}(q_f,\ell,s)\right) \\
&\qquad \times\left[1-n^+_F(\omega_{q,k},\Phi,\bar{\Phi})-n^-_F(\omega_{q,k},\Phi,\bar{\Phi})
\right]\;,
\label{flowu}
\end{align}
where 
$\omega_{1,k}=\sqrt{k^2+U_k^{\prime}}\,$, with $U_k^{\prime}=\partial U_k/\partial\rho$,
$\omega_{2,k}=\sqrt{k^2+U_k^{\prime}+2U_k^{\prime\prime}\rho}\,$,
$\omega_{q,k}=\sqrt{k^2+2g^2\rho}\,$,
$p^2_{\perp}(q,m,s)=(2\ell+1-s)|qB|\,$,
$n_B(x)=1/(e^{\beta x}-1)\,$ and 
$n_F^{\pm}(\omega_{q,k},\Phi,\bar{\Phi})$ are the generalised Fermi-Dirac distribution functions defined in Eqs.~(\ref{n1}) and (\ref{n2}) with $E=\omega_{q,k}$. As noted in~\cite{andersen12} there are two main methods used to solve this equation. One possibility is to use a polynomial expansion around the $k$-dependent minimum. The expansion is truncated and the coupled equations for the coefficients of the terms in the series are solved, see~\cite{kamikado13b} for a calculation using this method. Here instead we discretise all of the variables $(\rho,\Phi,\bar{\Phi},\,T,\mu,B)$, and solve the equation using standard numerical methods (described below). This is done in C++, although we calculate the gluonic potential and find the transition temperatures $T_c$ and $T_d$ using MatLab.

To find the expectation value of our three order paramters, $\sigma$, $\Phi$ and $\bar{\Phi}$ we evaluate $U_k$ on a grid in $(\rho,\Phi,\bar{\Phi})$-space, ultimately giving us $U_{k=0}(\rho,\Phi,\bar{\Phi})$. We then add to this result the gauge potential, $U_\trm{glue}(\Phi,\bar{\Phi})$ (which can be straightforwardly calculated given its simple analytic form). Finally in the minimisation procedure we must re-introduce the explicit symmetry breaking parameter described in Sec.~\ref{sec:RG}.
Thus finding $\langle\sigma\rangle$, $\langle\Phi\rangle$ and $\langle\bar{\Phi}\rangle$ we find the value of $\phi=\sqrt{2\rho}$, $\Phi$ and $\bar{\Phi}$ that minimise $U_{k=0} + U_\trm{glue} + h\phi$. This procedure can be then repeated for different values of $T$, $B$ and $\mu$ to build up the phase diagram.

%
\section{Evaluation of $U_{k=0}(\rho,\Phi,\bar{\Phi})$}
\label{sec:eval}

The major difficulty in the above (other than the sheer size of the phase space we examine) is the evaluation of $U_{k=0}(\rho,\Phi,\bar{\Phi})$. We use a grid in $\rho$ from 0 to 8000 MeV$^2$ in 200 (evenly spaced) steps, corresponding roughly to the region 0 to 126 MeV in $\sigma$. Note that this implies that the spacing of the steps in $\sigma$ is not even, with greater step density at higher $\sigma$.

We then step through the differential equation using a fourth order Runga-Kutta method. Each step requires the evaluation of first and second order derivative of $U_k$, along with a number of different sums over Landau levels for the various charged particles.\footnote{The sums over Landau levels are done by brute force. Although note that each sum is truncated by a term of the form $\theta(k^2-(2\ell+1)|qB|)$, thus this is only numerically intensive for small ($\lesssim 0.5$ $m_\pi$) magnetic field strengths.} We define a dimensionless RG time, $t=\trm{log}(k/\Lambda_\trm{RG})$, with $\Lambda_\trm{RG}=500$ MeV, and use this as the controlling parameter for the RG flow. We run from $t=0$ to $t=-6$, corresponding to $\Lambda_\trm{RG}=500$ MeV to $\Lambda_\trm{RG}\approx1$ MeV. At this point the position of the minimum of the potential is stable in RG time. It was shown in~\cite{schaefer05} that by running to $t=-\infty$ ($k=0$ MeV) the potential actually becomes convex, and the minimum is thus no longer uniquely defined.

\subsection{Complex nature of the potential}

Due to the derivative term in the bosonic frequencies (for example $\omega_{1,k}=\sqrt{k^2+U_k^{\prime}}\,$) $U_k$ becomes complex (although the boundary condition is real). This has two major implications, first we must decide what to do with the complex part of $U_{k=0}$. $U_{k=0}$ is a potential energy and thus $\trm{Re}[U_{k=0}]$ is simply the energy of that particular state, the imaginary part is then interpreted as the decay rate of the state, see~\cite{weinberg87}. As such we ignore the imaginary part in the minimisation procedure above, although we include it completely when calculating $U_{k=0}$. In addition we encountered numerical problems with the continuity of the phase of the complex number, and had to explicitly control this at each step of the integration.

\subsection{Adaptive Runga-Kutta, noise and numerical differentiation}

The keen reader will notice that the derivative terms in Eq.~(\ref{flowu}) not only give complex bosonic frequencies, but at some points we obtain $\omega_{k,1/2}=0$ resulting in undefined (infinite) points in $\partial_kU_k$. Due to the numerical integration, this ultimately leads to `noise' in the final result for $U_{k=0}$, although this noise is approximately constrained to the part of the potential that lies to the left of the minimum (as in this region $U_k^{\prime}<0$). Although these points (where $\omega_{k,1/2}=0$) are only created to the left of the minimum, in taking the derivative of the potential, we use the sixth order finite difference method to calculate $U_k^{\prime}$ and $U_k^{\prime\prime}$, which necessarily links the values of $U_k(\rho_i)$ to those between $\rho_{i-3}$ and $\rho_{i+3}$. In later steps of the integration these errors are further linked to surrounding points. In this way these errors can have some effect upon the minimum of the potential.
\begin{figure}[htbp]
	\centering
	\hspace{0in}
 	\includegraphics[width=3.5in]{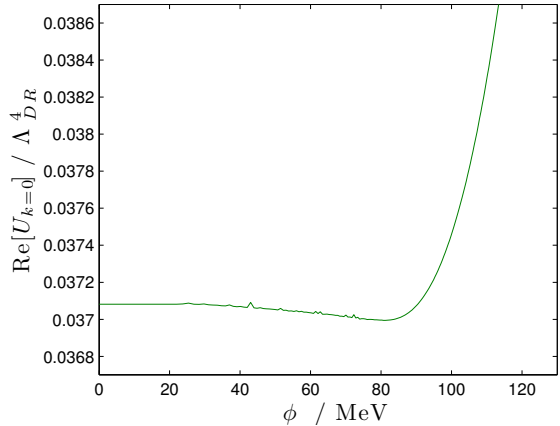}
	\caption[contentsCaption]{$U_{k=0}$ for $T=40$ MeV, $\mu=100$ MeV, $B=0$, $\Phi=\bar{\Phi}=1$ and at the physical point. We see visible noise for $\phi\lesssim85$ MeV originating from points in the flow where $\omega_{k,1}$ or $\omega_{k,2}$ are zero.}
	\label{potty}
\end{figure}
Figure~\ref{potty} gives $U_{k=0}$ for a point in the centre of the phase diagram ($T=40$ MeV, $\mu=100$ MeV, $B=0$, $\Phi=\bar{\Phi}=1$) below the chiral transition temperature and at the physical point (i.e.\ using physical masses for the pions), it thus represents a fairly typical curve for the potential (where chiral symmetry is broken). There exist, however, some parts of the phase diagram (for example in the chiral limit and with $\mu=0$), where noise is reduced to the point of being completely unnoticeable on a plot such as Fig.~\ref{potty}. Varying the input values of the Polyakov loop or the magnetic field seems not to have any great effect upon the relative strength of this noise.

Due to this noise it becomes both time consuming and inaccurate to use an adaptive Runga-Kutta method for solving the differential equation. The step size can decrease drastically even with fairly large error tolerances when encountering these points, however the quality (judged by the relative levels of noise) of the output $U_{k=0}$ is usually much worse. Instead we use a pre-set step size that increases as we step through the integration. The results are checked by comparison to existing calculations~\cite{schaefer05,andersen12} for a limited range of the phase space, to calculating using significantly smaller step size, and to an independent program written to solve the equation for a slightly expanded initial Lagrangian.

We are essentially concerned with the expectation value of our order parameters, and thus only in the minimum of $U_{k=0} + U_\trm{glue} + h\phi$. The combined effect of $U_\trm{glue} + h\phi$ is to push the minimum to slightly higher $\phi$ values, thus moving the point of interest further away from the region where the noise would affect our results. In the plots given in paper~\cite{wn1} the effects of this noise are insignificant as compared to the errors resulting from the procedure we use to find the transition temperatures (see Sec.~\ref{sec:ptn}).

We must take into account the numerical derivative when we choose the step spacing $\Delta\rho=\rho_{i+1}-\rho_i$. If this is chosen too small, then in the finite difference derivative we end up taking the difference of numbers that are very similar, and dividing by (in our case the seventh power of) a small number. We use 200 points ($\Delta\rho = 40$ MeV), as we see diminishing returns in terms of accuracy as we increase above this number.

\subsection{Polyakov loop order parameters}

As noted in~\cite{wn1} the surface $U_{k=0}(\Phi,\bar{\Phi})$ is extremely smooth, and as such we construct this via interpolation from an 8$\times$8 grid in $\Phi\times\bar{\Phi}$-space, with $\Phi$,~$\bar{\Phi} \in [0,1]$. We checked this interpolation against additional data at various points in the phase diagram and found negligible errors (on the order 0.1\%).


%
\section{Finding the critical temperatures}
\label{sec:ptn}

Working in the chiral limit ($h=m_\pi=0$) the chiral phase transition is a first order transition at low $T$ and high $\mu$ and second order elsewhere (see Fig.~1 of~\cite{schaefer05}). Thus finding the critical temperature $T_c$ amounts to finding the lowest temperature for which $\langle\sigma\rangle=0$. This can be done numerically very easily, furthermore, we can use previously known points in the phase diagram to inform later ones to quickly build up the entire phase diagram.

At the physical point the chiral phase transition is a cross-over transition, like that of Fig.~\ref{fig:ptd}. In this case we would ideally use the inflection point of the curve as the definition of the critical temperature, but doing this directly from the data would require a step size in $\rho$ that is unpractical using our numerical method. Thus we must fit the function first, and then find the inflection point the fitted function. As a fitting function we use
\begin{equation}
	a + b \left(\frac{\pi}{2} - \trm{arcTan}[c(T-d)]\right) \ep,
\end{equation}
where $a$, $b$, $c$ and $d$ are fit parameters. In~\cite{andersen12} a different fitting function was used, however we found this less effective at high $\mu$.

For the deconfinement transition coupled to the matter sector we always have a cross-over transition. However due to a change in the shape of both $\Phi(T)$ and $\bar{\Phi}(T)$ we are not able to find a single function that can suitably fit the transition for all values of $\mu$. As such, we interpolate $\Phi$ and $\bar{\Phi}$ in $T$ and find the point at which these interpolations are equal to $1/2$.



\appendix
%
\chapter{Appendix}
\label{cha:app}

\settocdepth{chapter}

%
\section{Useful sources and references}
\label{sec:ref}

For the interested student, and in gratitude to those that have provided great material from which I can work let me elaborate on some possible further reading.

First, most of this thesis was written with the theses of the former members of Jens' group, Lars K.~\cite{kyllingstad11}, Lars L.~\cite{leganger11} and Rashid K.~\cite{khan12} along with Peskin and Schroeder~\cite{peskin95}, Zee~\cite{zee10} and Maggiore's modern introduction~\cite{maggiore05} at hand. I highly recommend the chapters 2 and 3 of \cite{maggiore05} as a strong theoretical/group theoretical basis for the study of QFT. In addition Lars K.'s master's thesis~\cite{kyllingstad07} is accurately written and the Preliminaries section provides a useful reference and the section on the NJL model is far beyond what is presented here.

For paper~\cite{wn1} the RG calculations of Schaefer et al.~\cite{schaefer05,herbst10,herbst13} seem like foundational material. In addition Skokov's~\cite{skokov12} and Andersen and Traberg's~\cite{andersen12} works are clear precursors to our paper. The work of Kamikado and Kanazawa~\cite{kamikado13b} is also well worth reading for a treatment beyond the LPA.

For paper~\cite{wn2} I recommend only Fraga's paper~\cite{fraga13} although~\cite{andersen11} may prove helpful as a reference.

For paper~\cite{wn3} the lattice papers of Boz, Cotter and co-authors are essential reading~\cite{cotter13,boz13}. Additionally the following 2cQCD papers were of great help to me~\cite{andersen10,amador13,strodthoff14}.


  \settocdepth{section}

  \endgroup


 \introlayout

%

\nocite{*}
\bibliographystyle{unsrt}
\bibliography{main}



\part*{
  \normalfont\sffamily\bfseries\centering\Huge Paper 1\\
  \vspace{1ex}
  \titlerule
  \vspace{2ex}
  \normalsize\normalfont \newblock Chiral and deconfinement transitions in a magnetic background using the functional renormalization group with the Polyakov loop.\\
  \newblock {\em JHEP}, 1404:187, 2014.
  \thispagestyle{empty}}
  \includepdf[pages=-]{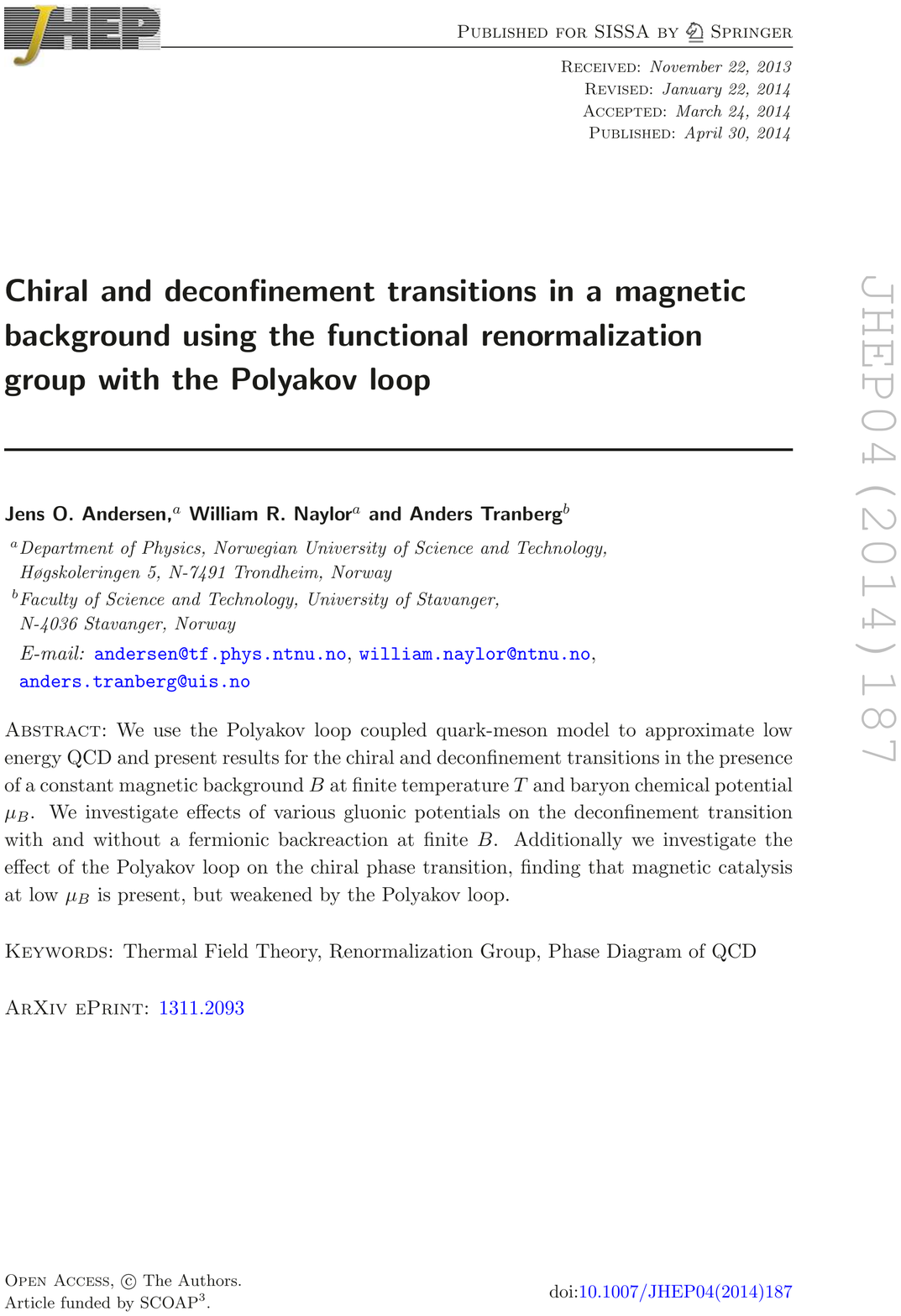}

\part*{
  \normalfont\sffamily\bfseries\centering\Huge Paper 2\\
  \vspace{1ex}
  \titlerule
  \vspace{2ex}
  \normalsize\normalfont \newblock Inverse magnetic catalysis and regularization in the quark-meson model.\\
  \newblock {\em JHEP}, 1502:205, 2015.
  \thispagestyle{empty}}
  \includepdf[pages=-]{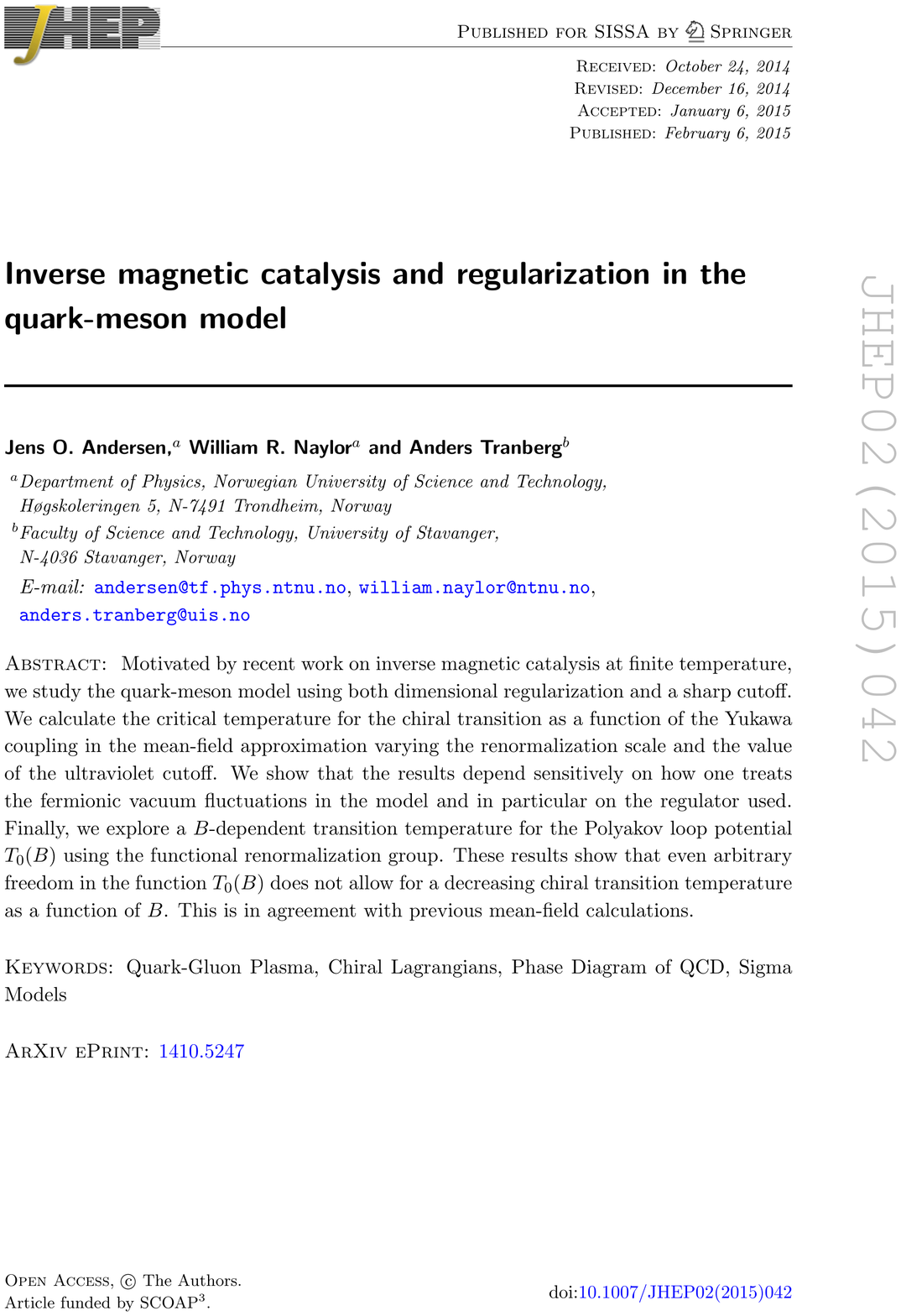}
  
\part*{
  \normalfont\sffamily\bfseries\centering\Huge Paper 3\\
  \vspace{1ex}
  \titlerule
  \vspace{2ex}
  \normalsize\normalfont \newblock Confronting effective models for deconfinement in dense quark matter with lattice data.\\
  \newblock Submitted to Physical Review D {\em arXiv:1505.05925}, 2015.
  \thispagestyle{empty}}
  \includepdf[pages=-]{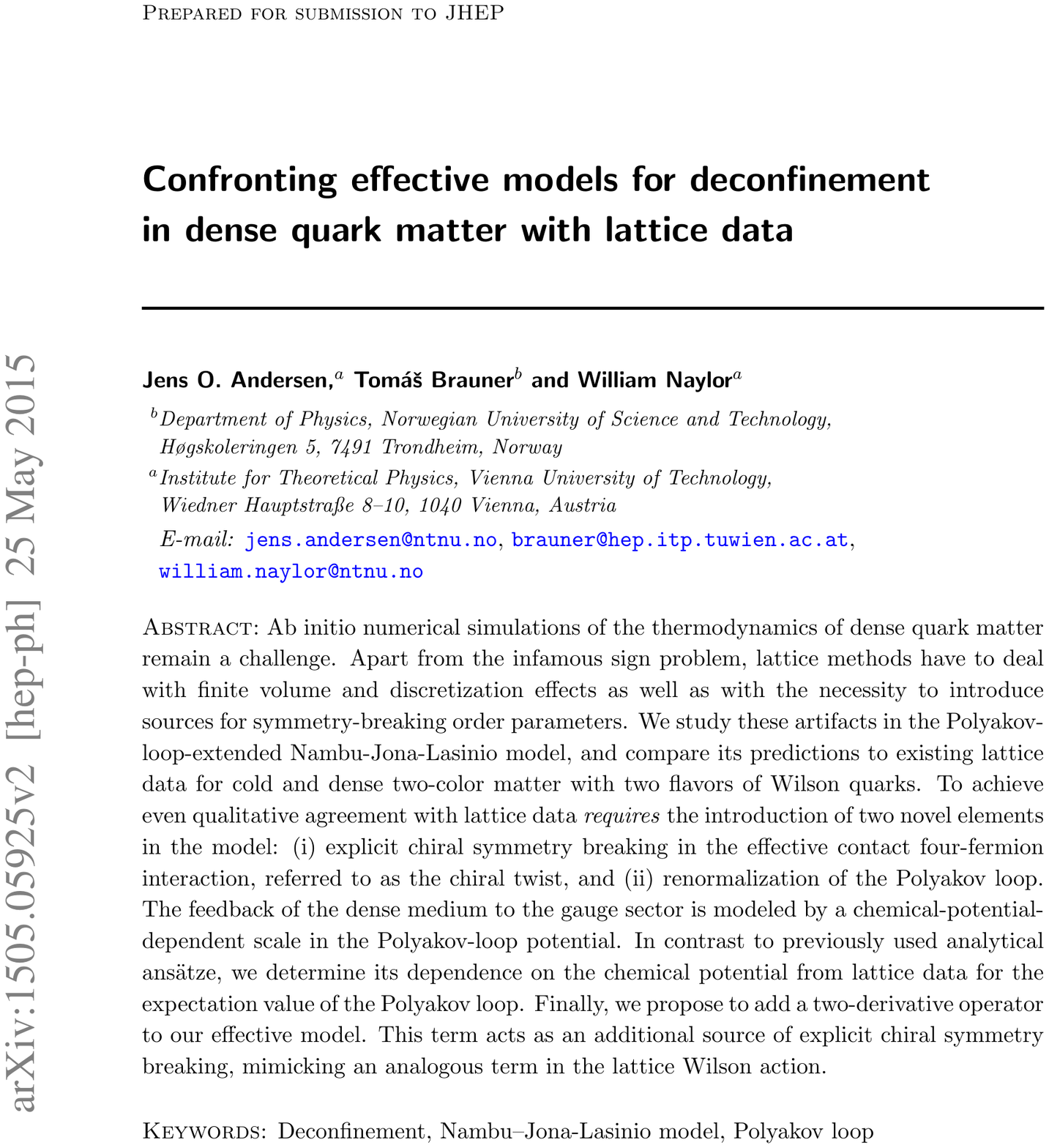}


\end{document}